\documentclass[useAMS,usenatbib]{mn2e}
\usepackage{graphicx}
\usepackage{subfigure}
\usepackage{epstopdf}
\usepackage{multirow}
\usepackage{amssymb}
\usepackage{amsmath}
\usepackage{float}

\newcommand{\mH}{{\rm H}}
\newcommand{\mHt}{{\rm H_{2}}}
\newcommand{\me}{{\rm e^{-}}}
\newcommand{\Hp}{{\rm H^{+}}}

\title[Cloud formation in colliding flows]{Cloud formation in colliding flows: influence of the choice of cooling function}
\author[Micic et al.]{Milica Micic$^{1,2,4}$\thanks{E-mail:milica@aob.rs}, Simon C. O. Glover$^{1}$, Robi Banerjee$^{3}$ and Ralf S. Klessen$^{1}$\\
$^{1}$Universit\"{a}t Heidelberg, Zentrum f\"{u}r Astronomie, Institut f\"{u}r Theoretische Astrophysik, Albert-Ueberle-Str. 2, 69120 Heidelberg\\
$^{2}$Member of the International Max Planck Research School for Astronomy and Cosmic Physics at the University of Heidelberg \\
(IMPRS-HD) and the Heidelberg Graduate School of Fundamental Physics (HGSFP)\\
$^{3}$ Hamburger Sternwarte, Gojenbergsweg 112, 21029 Hamburg, Germany \\
$^{4}$ Astronomical Observatory, Volgina 7, 11060 Belgrade, Republic of Serbia \\ 
}

\begin{document}


\pagerange{\pageref{firstpage}--\pageref{lastpage}} \pubyear{2012}

\maketitle

\label{firstpage}

\begin{abstract}

We study the influence of the choice of cooling function on the formation of molecular clouds in high-resolution three-dimensional simulations of converging flows. We directly compare the results obtained using the simple, parametrized cooling function introduced by \citet{ki02} and used by a number of converging flow studies with the results of the detailed calculation of the non-equilibrium chemistry and thermal balance of the gas. We find that a number of the cloud properties, such as the mass and volume filling fractions of cold gas, are relatively insensitive to the choice of cooling function. On the other hand, the cloud morphology and the large-scale velocity distribution of the gas do strongly depend on the cooling function. We show that the differences that we see can largely be explained by differences in the way that Lyman-$\alpha$ cooling is treated in the two complementary approaches, and that a proper non-equilibrium treatment of the ionisation and recombination of the gas is necessary in order to model the high-temperature cooling correctly.

We also investigate the properties of the dense clumps formed within the cloud. In agreement with previous models, we find that the majority of these clumps are not self-gravitating, suggesting that some form of large-scale collapse of the cloud may be required in order to produce gravitationally unstable clumps and hence stars. Overall, the physical properties of the dense clumps are similar in both simulations, suggesting that they do not depend strongly on the choice of cooling function. However, we do find a systematic difference of around 10~K in the mean temperatures of the clumps produced by the two models.

\end{abstract}


\section{Introduction}

An important goal of star formation research is to understand the formation and evolution of giant molecular clouds (GMCs), known to be the hosts of all observed Galactic star formation. However, as yet this remains an unsolved problem. Traditionally, molecular clouds were viewed as being virialized structures in the interstellar medium (ISM), having relatively long lifetimes and a significant delay between the formation of the cloud and the onset of star formation \citep[see e.g.][]{zp74,wood78,sh79}. However, recent studies have suggested that GMCs begin forming stars shortly after they themselves form, and are non-equilibrium entities \cite[see e.g.][]{bp99a,bp99b,el00,hartmannetal01,vaz03,vaz06}. Their formation and evolution are dominated by the effects of supersonic turbulent motions (see e.g.\ \citealt{lar81,my83,sol87,fal92,hb04} and the reviews by \citealt{rk04,es04,se04,mo07}) and are rapid, with a timescale comparable to those of important chemical processes such as the conversion of atomic to molecular hydrogen or the freeze-out of molecules onto the surfaces of interstellar dust grains. Therefore, the dynamics and chemistry of the gas are strongly coupled, with one directly influencing the evolution of the other, and they must be modelled together. 

A promising theory for producing GMCs, as well as for generating turbulence within them, suggests that these clouds form in places where streams of warm atomic gas collide. Work by a number of different groups has shown that the dense sheets and filaments that build up at the interface of the colliding flows become thermally unstable \cite[e.g.][]{audit05,vaz07,hen08,robi09,heitsch08}. The compressed unstable gas rapidly radiates away most of its thermal energy, significantly decreasing its temperature. As the temperature of the gas falls, it is compressed by the warmer material surrounding it, and so the large drop in temperature is associated with a large increase in density. In some regions, strong pressure gradients are created that then act to drive turbulent flows throughout the interaction region. The outcome is a set of dense, cold clumps of gas, embedded in a more diffuse, turbulent flow, and with a filamentary morphology reminiscent of that found by observations of real GMCs \cite[e.g.][]{men10,arz11}. Since the dense clumps observed in real GMCs are known to be the sites at which stars form, these results suggest that there is an important link between the large scale physics of GMC formation and the smaller scale physics of star formation.

Unfortunately, most three-dimensional studies of this process performed so far make use of highly simplified treatments of the thermal energy balance of the gas. Heating and cooling are modelled using simplified parameterizations that assume that the rates are functions only of density and temperature, and that ignore the effects of chemical changes or dust shielding \cite[see e.g.][]{vaz07,robi09}. No effort has been made to determine whether this simplified approach is adequate for describing the dynamics of the gas, or whether we would expect to find significant changes in behaviour if we were to use a more detailed chemistry and cooling model.  

In this paper, we attempt to improve on this situation. We perform high-resolution 3D simulations of cloud formation in colliding flows, and directly compare the results obtained from a simplified cooling model with those that we obtain from a self-consistent treatment of the cooling and chemistry of the gas. Our main goal is to understand how the use of a more accurate thermal model affects the dynamics of the flow and the nature of the structures that form within it.

\section{Numerical Model}

\subsection{The numerical code and setup}
\label{setup}
We consider the atomic phase of the interstellar medium (ISM), whose behaviour is governed by the equations
\begin{equation}
\frac{\partial\rho}{\partial t}+\nabla\cdot(\rho \boldsymbol\upsilon)=0
\end{equation}
\begin{equation}
\frac{\partial(\rho \boldsymbol\upsilon)}{\partial t}+\nabla\cdot(\rho \boldsymbol{\upsilon\upsilon})=-\nabla P + \frac{1}{4\pi}\left( \nabla \times \mathbf{B} \right) \times \mathbf{B},
\end{equation}
\begin{equation}
\frac{\partial E}{\partial t}+\nabla\cdot[(E+P)\boldsymbol\upsilon]=n \Gamma-n^{2}\Lambda(T),
\end{equation}
\begin{equation}
\frac{\partial \mathbf{B}}{\partial t} = \nabla \times \left(\boldsymbol\upsilon \times \mathbf{B} \right),
\end{equation}
where $\rho$ is the gas density, $\boldsymbol\upsilon$ is the fluid velocity, $\mathbf{B}$ is the magnetic field strength, which also must satisfy the constraint $\nabla \cdot \mathbf{B} = 0$,
$E=P/(\gamma-1)+\rho|v|^{2}/2$ is the total energy per unit volume, $P$ is the thermal pressure, and $\gamma=5/3$ is the adiabatic index. In the energy equation, $n$ is the number density of hydrogen nuclei, which is related to the mass density via $n=\rho/(1.4 m_{\rm p})$, where $m_{\rm p}$ is the proton mass, $n \Gamma$ is the radiative heating rate per unit volume and $n^{2} \Lambda$ is the radiative cooling rate per unit volume. In writing down this set of equations, we have assumed that we can neglect the effects of the self-gravity of the gas. We make a similar assumption in the simulations presented in this paper, and defer investigation of the more computationally demanding self-gravitating case to future work.

We model the collision of two large cylindrical streams of warm atomic gas using a modified version of the adaptive mesh refinement (AMR) code \textsc{FLASH} \citep[]{fr00}. Our modifications include the addition of a simplified but accurate treatment of the most important hydrogen chemistry, together with a detailed atomic and molecular cooling function. They are described in detail in \citet[]{mm12}. For our study we use the MHD 5-wave Bouchut solver which preserves positive states for the density and pressure \citep[]{bkw07,bkw10,waagan09,wfk11}. This implementation also applies the truncation-error method \citep[]{powell99} to maintain small $\nabla \cdot \mathbf{B}$ errors and to avoid unphysical magnetic tension terms.

We use a similar setup to that studied in \citet{robi09}, which itself was based on model L256$\Delta v$0.17 from the study of \citet{vaz07}. The two cylindrical streams, each 112 pc long and 32 pc in radius, are given an initial, slightly supersonic inflow velocity so that they collide at the centre of the numerical box ($x=0$ pc). The (256~pc)$^{3}$ simulation box is periodic, and the streams are completely contained within it, such that the resulting cloud occupies a relatively small volume far from the boundaries, and interacts freely with its diffuse environment, with relatively little effect from the boundaries. The box is initially filled with warm atomic gas with a uniform number density $n$ = 1 cm$^{-3}$. This corresponds to an initial mass density of
$\rho$ = 2.12$\times10^{-24}$ g cm$^{-3}$, if we adopt a 10:1 ratio of hydrogen to helium, by number. The initial temperature of the atomic gas is $\sim5000$K, corresponding to an isothermal sound speed of 5.7 km s$^{-1}$. The initial velocity of each flow is $7 \: {\rm km} \: {\rm s^{-1}}$, and so the flows have an initial isothermal Mach number of 1.22. At temperature of $T=5000$K for the warm phase, this implies that the cold phase comes into hydrostatic thermal pressure balance with the warm gas at a density of roughly 100 cm$^{-3}$ \citep[see e.g.][]{w95,w03}. Furthermore, we add 10\% random velocity perturbations to the bulk stream. Finally, we note that we include a magnetic field, which we assume to be oriented parallel to the inflow. The initial magnetic field strength is taken to be $3 \: \mu$G, consistent with estimates of the mean Galactic magnetic field strength \citep{beck01}. This corresponds to a critical mass-to-flux ratio \citep[see][]{vaz11}.

We follow the collision of the streams of gas with up to 11 AMR refinement levels corresponding to a maximum effective resolution of $8192^{3}$ grid cells, or a grid spacing of $\Delta x$ = 0.03~pc in each direction. We use a Jeans-type criterion \citep[]{truelove97, fed11} for the dynamical mesh refinement which requires that the local Jeans length is resolved with at least 10 grid cells while refinement is active. 
Our use of a Jeans-type criterion ensures that we resolve any structures
large enough to be gravitationally unstable (although as we do not
include the effects of self-gravity in these simulations, these
structures will not actually undergo gravitational collapse). In practice,
we find that this criterion also allows us to resolve a considerable
amount of the smaller-scale structure of the gas, as discussed in more
detail in Section 3. Nevertheless, our limited resolution means that
we will still miss structures on very small scales, as we expect the
thermal instability to create structures on all scales larger than the
Field length \citep{field65}, which in the CNM is typically of the order
of $10^{-3} \: {\rm pc}$ or less \citep[see e.g.][]{gml07a}.
Fortunately, previous studies have shown that the failure to resolve
this very small scale structure does not significantly affect the
dynamics of the gas on larger scales \citep[see e.g.][]{gressel09}, and
hence this limitation should not significantly affect our results.

\subsection{Cooling and heating}
\label{chemcool}

The simulations of \citet[]{vaz07} and \citet[]{robi09} used a cooling function derived from the one-dimensional colliding flow models of \citet{ki00}. A simple analytical fit was provided by \citet[]{ki02}, and has the form\footnote{Note that the version of this fit printed in \citet{ki02} suffers from a significant typographical error, which has the effect of
making the low-temperature cooling rate far too large. The version we quote here is the corrected version of their fit, as given in \citet{vaz07}.}: 
\begin{equation}
\Gamma=2.0\times10^{-26}~\rm{erg}~\rm{s^{-1}},
\end{equation}
\begin{equation}
\begin{split}
\frac{\Lambda(T)}{\Gamma}=10^{7} \exp \left(-\frac{1.184\times10^{5}}{T+1000}\right) \\
+1.4\times10^{-2}\sqrt{T} \exp \left(-\frac{92}{T}\right)~ \rm{cm}^{3}.
\end{split}
\end{equation}
where $T$ is the gas temperature in Kelvin. With this cooling function, the simulated ISM is thermally unstable in the density range $1\lesssim n\lesssim10$ cm$^{-3}$, corresponding to equilibrium temperatures in the range $500\lesssim T\lesssim 5000$~K. 

This simplified treatment of the heating and cooling of the gas does not account for any chemical effects. It assumes that the photoelectric heating efficiency is constant, whereas in practice it is known to depend on the electron number density \citep[see e.g.][]{bt94,w95}. It also assumes that Lyman-$\alpha$ cooling is extremely efficient: the high temperature cooling rate assumed in the \citet{ki02} function corresponds to the cooling one would expect in a gas in which the electron and atomic hydrogen number densities are approximately equal. In our present study, we investigate how the results that we obtain when we properly account for these effects compare with the results produced by this more simplified treatment.
\begin{table}
\caption{Reactions in our non-equilibrium chemical model. \label{chem_model}}
\begin{tabular}{rlc}
\hline
No.\ & Reaction & Reference \\
\hline
1 & $\mH + \mH + {\rm grain} \rightarrow \mHt + {\rm grain}$ & 1 \\
2 & $\mHt + \mH \rightarrow \mH + \mH + \mH$ & 2 \\
3 & $\mHt + \mHt \rightarrow \mH + \mH + \mHt$ & 3 \\
4 & $\mHt + \me \rightarrow \mH + \mH + \me$ & 4 \\
5 & $\mH + {\rm c.r.} \rightarrow \Hp + \me$ & See \S\ref{chemcool} \\ 
6 & $\mHt + {\rm c.r.} \rightarrow \mH + \mH$ & See \S\ref{chemcool} \\ 
7 & $\mHt + {\rm c.r.} \rightarrow \mH + \Hp + \me$ & See \S\ref{chemcool} \\
8 & $\mH + \gamma_{\rm X} \rightarrow \Hp + \me$ & 5 \\ 
9 & $\mH + \me \rightarrow \Hp + \me + \me$ & 6 \\  
10 & $\Hp + \me \rightarrow \mH + \gamma$ & 7 \\ 
11 & $\Hp + \me + {\rm grain} \rightarrow \mH + {\rm grain}$ & 8 \\ 
\hline
\end{tabular}
\medskip
\\
{\bf Notes}: ``c.r.'' denotes a cosmic ray particle, and $\gamma_{\rm X}$ denotes an X-ray photon \\
{\bf References}: 1: \citet{hm79}, 2: \citet{ms86}, 3: \citet{mkm98}, 
4: \citet{tt02}, 5: \citet{w95}, 6: \citet{a97}, 7: \citet{f92}, 8: \citet{wd01a}
\end{table}

To do this, we perform two simulations using the setup described in Section~\ref{setup}. In one, we use the \citet{ki02} cooling function, as in \citet{robi09}. In the other, we use the time-dependent chemical model and cooling function described in \citet{gml07a,gml07b}. This model follows the abundances of four chemical species -- free electrons, H$^{+}$, H, and H$_{2}$ -- linked by the reactions listed in Table~\ref{chem_model}. It assumes that any carbon in the gas remains in the form of C$^{+}$ and that any oxygen remains in atomic form, and so underestimates the cooling rate in regions dominated by CO. However, \citet{gc12} have shown that this does not have a strong effect on the dynamics of the gas on scales larger than those of individual pre-stellar cores, and so making this simplifying assumption should not greatly affect our results.

We use an implementation of the Glover \& Mac~Low model within \textsc{FLASH} that is described in detail in \citet{mm12}. Our approach uses \textsc{FLASH}'s standard tracer field implementation to directly follow the advection of the fractional abundances of molecular hydrogen ($x_{\rmn{H_{2}}}$) and ionised hydrogen ($x_{\rmn{H^{+}}}$). The abundances of the other two species -- atomic hydrogen ($x_{\rmn{H}}$) and electrons ($x_{\rmn{e}}$) -- are computed from the conservation laws 
for charge
\begin{equation}
x_{\rmn{e}}=x_{\rmn{H^{+}}}+x_{\rmn{C^{+}}}+x_{\rmn{Si^{+}}}
\end{equation}
and for the total amount of hydrogen
\begin{equation}
x_{\rmn{H}}=x_{\rmn{H,tot}}-x_{\rmn{H^{+}}}-2x_{\rmn{H_{2}}}.
\end{equation}
Here, $x_{\rmn{H,tot}}$ is the total abundance of hydrogen nuclei in all forms, which we normalize to unity, and $x_{\rmn{C^{+}}}$ and $x_{\rmn{Si^{+}}}$ are the abundances of ionised carbon and silicon, respectively, which remain fixed throughout the simulations.

\begin{figure*}[h]
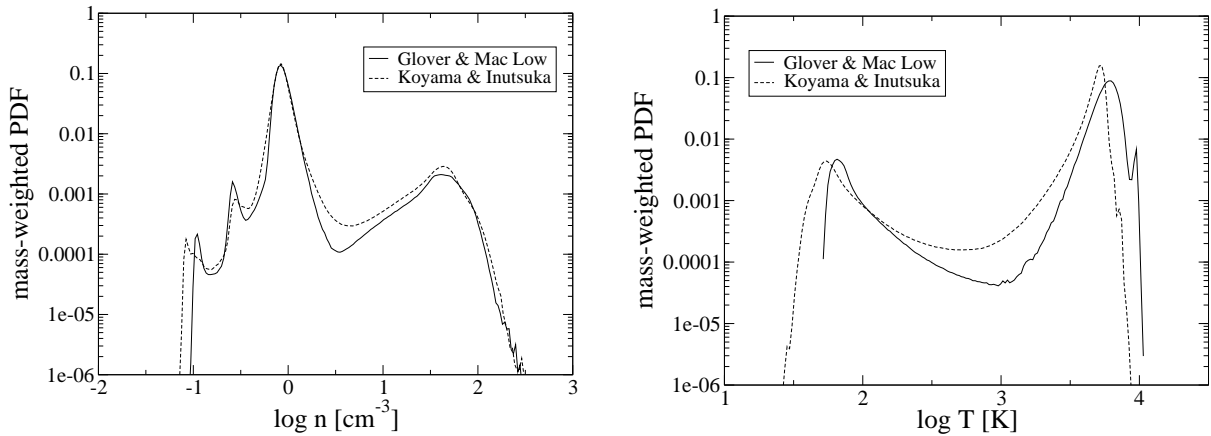

\centering
\includegraphics[width=7.6cm]{f01a.eps}
\qquad
\includegraphics[width=7.7cm]{f01b.eps} 
\caption{Mass-weighted density (left panel) and temperature (right panel) probability distribution function (PDF)
at time $t=22$ Myr. The solid line presents the PDF in the run with non-equilibrium chemical model, while the dashed line corresponds to the PDF in the run with Koyama \& Inutsuka cooling function.}
\label{fig:PDF}
\end{figure*}

We assume that carbon, oxygen and silicon remain in the form of C$^{+}$, O and Si$^{+}$ throughout the simulation, and adopt fractional abundances for these species given by $x_{\rmn{C^{+}}} = 1.41 \times 10^{-4}$, $x_{\rmn{O}} = 3.16 \times 10^{-4}$ and $x_{\rmn{Si^{+}}} = 1.5 \times 10^{-5}$, respectively \citep[]{s00}. The radiative and chemical heating and cooling of the gas is modelled with a cooling function that contains contributions from a range of processes, of which the most important are C$^{+}$ and O fine structure cooling, Lyman-$\alpha$ cooling, and photoelectric heating. Full details of these processes, along with the other contributions to our cooling function, can be found in \citet{gml07a,gml07b} and \citet{mm12}.

In our run with the Glover \& Mac~Low chemistry and cooling model, we adopt a value of $G_{0} = 1.0$ for the strength of the interstellar radiation field in units of the Habing field \citep{habing68}. We take the cosmic ray ionisation rate of atomic hydrogen to be $\zeta_{\rm H} = 10^{-17} \: {\rm s^{-1}}$ and assume that the ratio of this rate to the cosmic ray ionisation rate of H$_{2}$ is the same as given in the UMIST astrochemistry database \citep{umist99}. We include the effects of X-ray ionisation and heating using the prescription given in Appendix A of \citet{w95}, and assume a uniform absorbing column density of warm atomic hydrogen $N_{\rm w} = 10^{19} \: {\rm cm^{-3}}$. In our present study, we do not include the effects of self-gravity, dust shielding, or H$_{2}$ self-shielding. We note that although we expect dust shielding to have a significant effect on the thermal state of the gas regions with mean visual extinctions $\bar{A}_{V} > 1$, these account for only a small fraction of the simulation volume. 

\section{Results}
\subsection{Density and temperature distributions}
\label{dense}

The sequence of events that occurs within our two simulations is broadly similar in both cases, and is also in good agreement with the results of previous studies \citep[see e.g.][]{vaz07,robi09}. We therefore begin by briefly describing this sequence of events, before moving on to look at the differences that do occur between the two runs.

At the interface where our transonic, converging flows collide, the gas is shocked and moderately compressed. This compression destabilises the gas, triggering a thermal instability (TI) that causes the gas to cool rapidly. As the gas cools, it is compressed by the thermal pressure of the surrounding warm gas, leading to a rapid increase in its density. This process comes to an end once the gas reaches the equilibrium temperature of the cold neutral medium (CNM) phase, which for the conditions simulated here is below 100~K. The cool dense gas initially forms a sheet that then fragments into filaments and ultimately into small, pressure-confined clumps. As the thermal pressure of the dense gas is in close balance with the total (thermal plus ram) pressure of the warm neutral medium (WNM) outside it, the cold gas can easily reach number densities of the order of $100 \: {\rm cm^{-3}}$, comparable to the mean density of the gas in many GMCs \citep[see e.g.][]{rd10,hughes10}. 

The cloud of gas that forms in the interface region is composed of a mixture of diffuse and dense gas, including a significant fraction of material in the thermally unstable region intermediate between the CNM and WNM phases (see Figure~\ref{fig:PDF}). Rather than the classical picture of a two-phase medium, we find instead a continuous distribution of densities and temperatures, albeit one with clear peaks corresponding to the CNM and WNM regimes \citep[see also][]{vaz00,gaz01,gaz05,audit05}. 

\begin{figure*}[h]
\centering
\includegraphics[width=7cm]{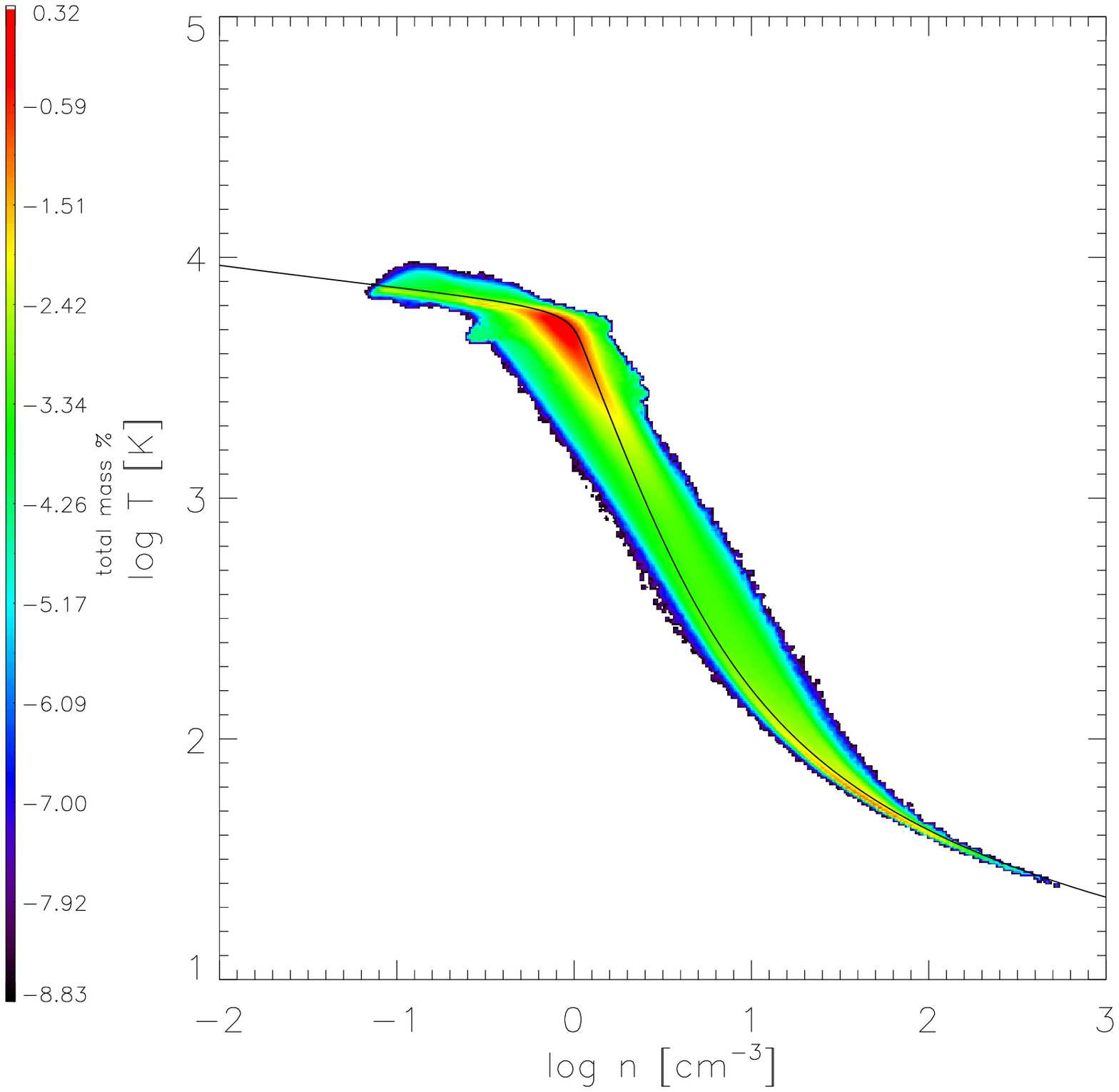}
\qquad
\includegraphics[width=7cm]{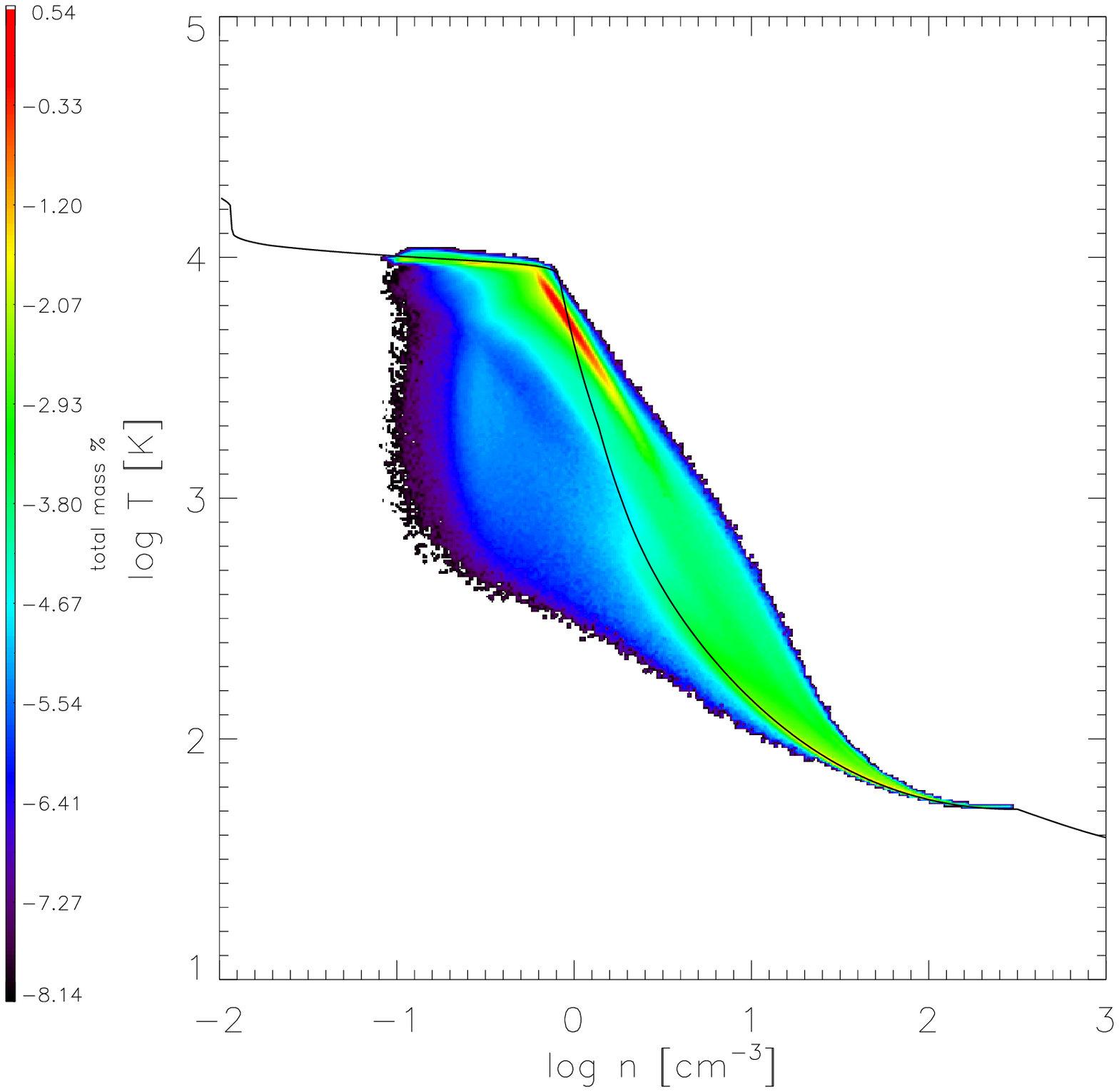}
\caption{Two-dimensional PDFs of temperature and density for the two simulations, for a time $t = 22$~Myr. The 
left-hand panel shows the results from the simulation that used the Koyama \& Inutsuka cooling function, and the 
right-hand panel shows the results from the simulation using the full non-equilibrium treatment. The fraction of
mass in each region of the density-temperature space is indicated by the colour scale. The solid line shows the
equilibrium temperature as a function of density, derived under the assumption that the gas is also in chemical
equilibrium.}
\label{fig:temp_eq}
\end{figure*}

In Figure~\ref{fig:PDF} we plot mass-weighted density (left-hand panel) and temperature (right-hand panel) probability distribution functions (PDFs) for both models at a time $t=22$~Myr, several million years after the end of the inflow. It is clear from the Figure that these PDFs do not differ by much between the two runs. The main difference that is apparent is that the temperatures that we recover for the CNM and WNM phases (the two clear peaks in the temperature PDF) are slightly smaller in the simulation run using the Koyama \& Inutsuka cooling function than in the simulation using the Glover \& Mac~Low treatment. This difference in behaviour is relatively simple to understand. At high temperatures ($T > 7000$~K), the Koyama \& Inutsuka cooling rate coefficient is given approximately by
\begin{equation}
\Lambda_{\rm KI}(T) \simeq 2 \times 10^{-19} \exp \left(-\frac{1.184\times10^{5}}{T+1000}\right).
\end{equation}
The main coolant in the Glover \& Mac~Low treatment at these temperatures is Lyman-$\alpha$ cooling, for which 
they use the following expression from \citet{cen92}
\begin{equation}
\Lambda_{\rm Ly-\alpha} = \frac{7.5 \times 10^{-19}}{1 + \sqrt{T / 10^{5}}} \exp \left(- \frac{118348}{T} \right)
\: x_{\rm e} x_{\rm H},
\end{equation}
where $x_{\rm e} \equiv n_{\rm e} / n$ is the fractional abundance of electrons and $x_{\rm H} \equiv n_{\rm H} / n$ is the fractional abundance of atomic hydrogen. Comparing these two cooling rates, we find that they produce comparable amounts of cooling only when $x_{\rm e} \simeq x_{\rm H} \simeq 1/2$, i.e.\ only when the chemical state of the gas is such that we get roughly the maximum amount of Lyman-$\alpha$ cooling possible. In highly ionised gas with $n_{\rm e} \gg n_{\rm H}$, or predominantly neutral gas with $n_{\rm e} \ll n_{\rm H}$, the Lyman-$\alpha$ cooling rate is considerably smaller, and hence in these conditions, the Koyama \& Inutsuka treatment significantly overestimates the true cooling rate of the gas. The exponential temperature dependence of the cooling rate means that a large error in the value of the rate leads to only a small error in the gas temperature, but this is sufficient to explain the offset in the characteristic temperature of the WNM that we find when we compare our two simulations.

The difference in the CNM temperatures is not caused by a difference in the low temperature cooling rates, but rather by a difference in the radiative heating rate. In the Koyama \& Inutsuka treatment, the heating rate throughout the gas is simply $\Gamma = 2 \times 10^{-26} \:{\rm erg} \: {\rm s^{-1}}$. In the Glover \& Mac~Low treatment, on the other hand, the heating rate is sensitive to the chemical composition of the gas owing to the fact that the photoelectric heating efficiency is a function of the electron density \citep{bt94,wd01b}. In cold, dense gas, photoelectric heating is relatively efficient, and the heating rate is given approximately by $\Gamma_{\rm pe} \simeq 5 \times 10^{-26} \: {\rm erg} \: {\rm s^{-1}}$, i.e.\ it is roughly 2.5 times larger than assumed in the Koyama \& Inutsuka treatment. It is therefore not surprising that the gas can cool to somewhat lower temperatures in this case.

\begin{figure*}[h]
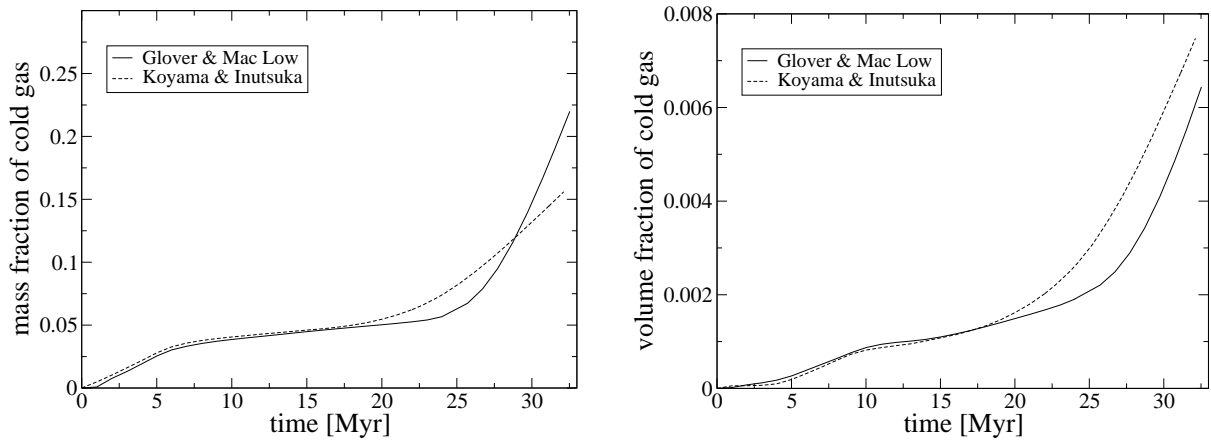

\centering
\includegraphics[width=7.6cm]{f03a.eps}
\qquad
\includegraphics[width=7.7cm]{f03b.eps}
\caption{Evolution with time in our two simulations of the mass (left-hand panel) and volume (right-hand panel) fractions 
of cold gas, defined here as gas with temperature $T\leq300$K. The solid line is for the run with the non-equilibrium 
chemical model, while the dashed line is for the run with the Koyama \& Inutsuka cooling function.}
\label{fig:cold_vol}
\end{figure*}

In order to investigate the dependence of temperature on density in our two simulations, we plot in Figure~\ref{fig:temp_eq} the two-dimensional (2D) PDFs of these quantities. The results for the Koyama \& Inutsuka run are shown in the left-hand panel, while the results of the non-equilibrium run are shown in the right-hand panel. The equilibrium temperature of the gas is indicated using the solid line. In the case of the simulation using the non-equilibrium treatment of chemistry and cooling, we derived an equilibrium temperature at each density by assuming that the gas was also in chemical equilibrium. We see from Figure~\ref{fig:temp_eq} that in the simulation with the Koyama \& Inutsuka cooling function, most of the gas has a temperature close to the equilibrium value. In the simulation with the Glover \& Mac~Low chemistry and cooling, on the other hand, the departures from equilibrium are more pronounced. The gas is close to the equilibrium temperature at densities $n < 1 \: {\rm cm^{-3}}$ and $n > 30 \: {\rm cm^{-3}}$, but the equilibrium temperature curve does not give a good description of the distribution of gas temperatures at intermediate densities. In this intermediate regime, the temperature falls off less rapidly with increasing density than predicted by the equilibrium temperature curve. This is a consequence of the sensitivity of the photoelectric heating rate to the electron number density, $n_{\rm e}$. As $n_{\rm e}$ increases, the net positive charge of the dust grain population decreases, or even becomes negative. This makes it easier for incoming photons to cause the ejection of photoelectrons. Consequently, the photoelectric heating efficiency tends to increase with increasing electron number density, up to a limiting value of a few percent \citep{bt94,w95,wd01b}. In the density and temperature regime where we see the greatest deviations from the equilibrium temperature curve, the photoelectric heating efficiency has not yet reached this limiting value, and hence any difference between the actual electron number density and the equilibrium value leads to a photoelectric heating rate that also differs from the value that it would have in equilibrium. As the recombination timescale in this gas is relatively long, of the order of a few Myr or more, the gas is generally slightly more ionised than it would be in equilibrium, and hence is heated slightly more efficiently. 

Finally, as we know that stars form in cold gas, it is interesting to examine how the cold gas fraction differs in our two simulations. As we have already seen, the cold and warm phases in our simulations are not completely distinct, and a significant fraction of the mass of the cloud lies intermediate between these two phases. The definition of "cold'' gas is therefore somewhat subjective. In this paper, we define cold gas to be gas with a temperature $T < 300$~K. (We note from Figure~\ref{fig:temp_eq} that in both simulations, the majority of the gas with a temperature as low as this has a density $n > 10 \: {\rm cm^{-3}}$). In Figure~\ref{fig:cold_vol}, we plot how the mass fraction (left-hand panel) and volume fraction (right-hand panel) of the cold gas evolves with time in both simulations.

Prior to the end of the inflow, at $t \sim 16$~Myr, both simulations show extremely similar behaviour. The amount of cold gas is small -- it accounts for only 5\% of the total mass and only a tiny fraction of the total volume of the simulation -- and the cold gas fraction increases only slowly with time. After the end of the inflow, however, greater differences become apparent between the two runs. In the run with the non-equilibrium cooling and chemistry, the cold gas fraction remains small up to $t \sim 25 \: {\rm Myr}$, but thereafter begins to increase rapidly. On the other hand, in the run with the Koyama \& Inutsuka cooling function, the cold gas fraction grows steadily from time $t \sim 20 \: {\rm Myr}$ until the end of the simulation, at a somewhat faster rate than during the inflow phase. In the interval $20 < t < 28$~Myr, there is more cold gas in the Koyama \& Inutsuka run than in the non-equilibrium run, but at later times the latter run has the most cold gas. Figure~\ref{fig:cold_vol} also indicates that the volume filling factor of the cold gas in the Koyama \& Inutsuka run is larger than in the other run, although the difference between the two is not large. We can understand the difference in the behaviour of the cold gas fraction in the two run by looking at how the velocity field generated in the cloud differs between the runs, which we examine in the next section.

\begin{figure*}
\centering
\includegraphics[width=5.68cm]{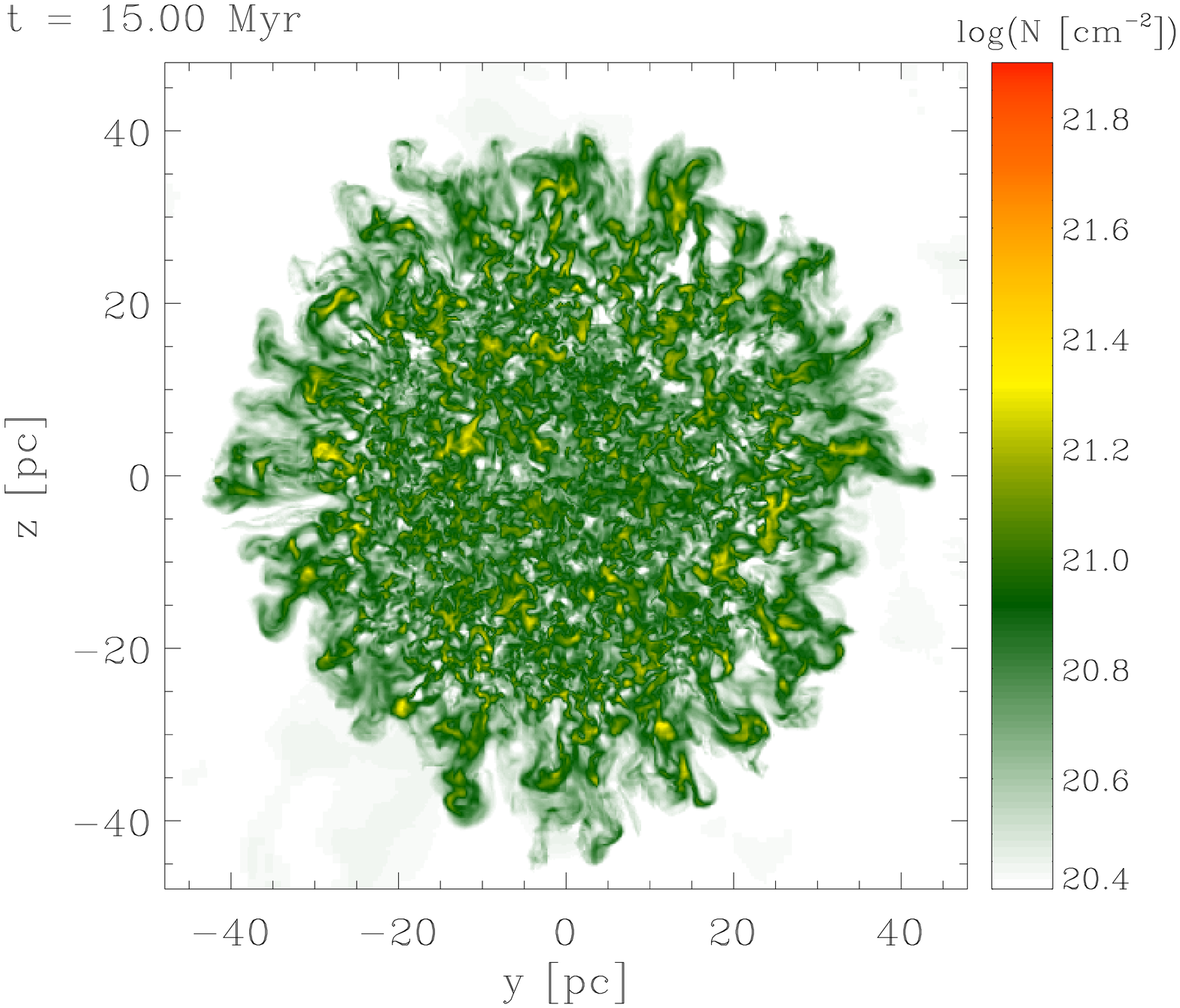}
\includegraphics[width=5.68cm]{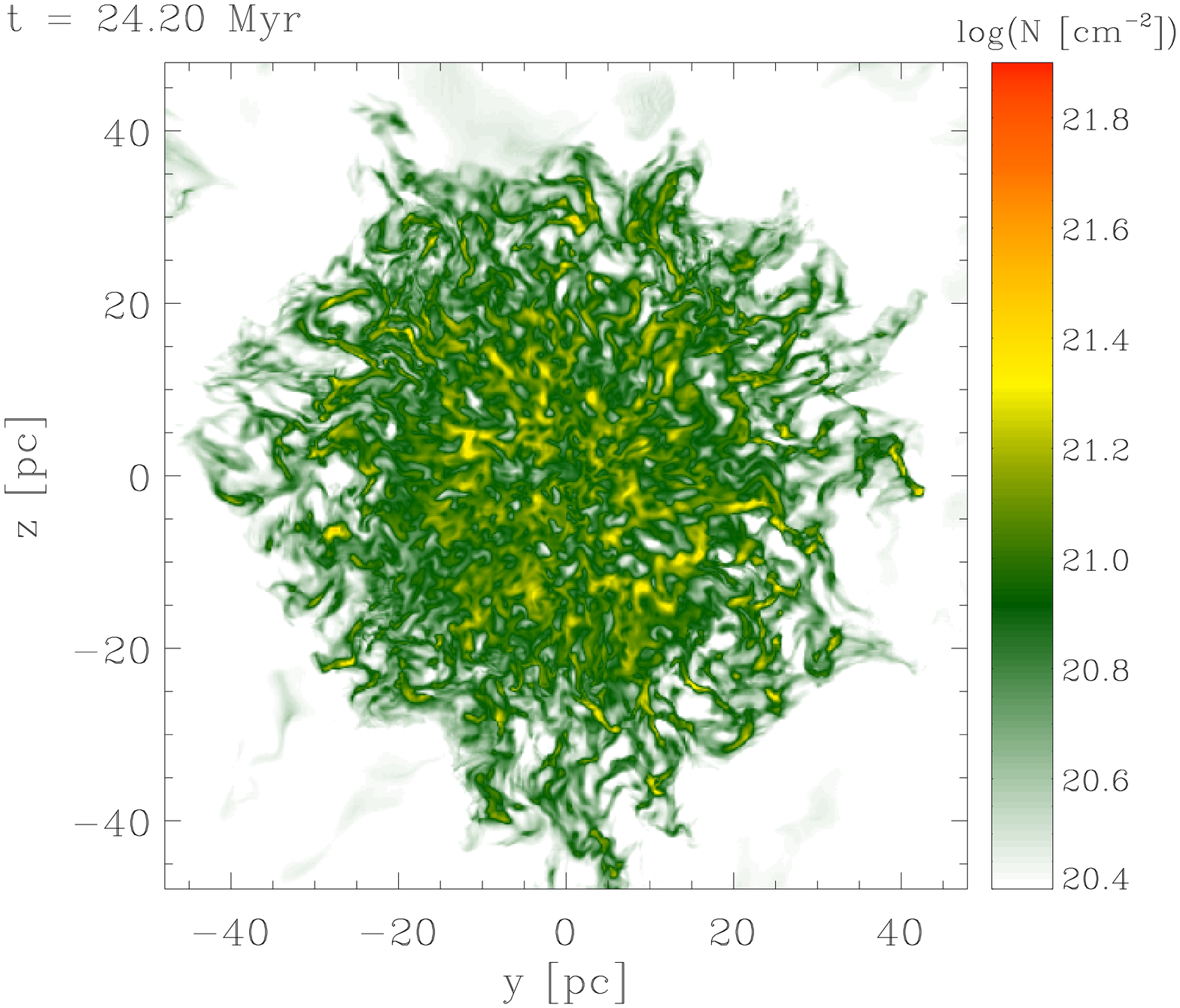}
\includegraphics[width=5.68cm]{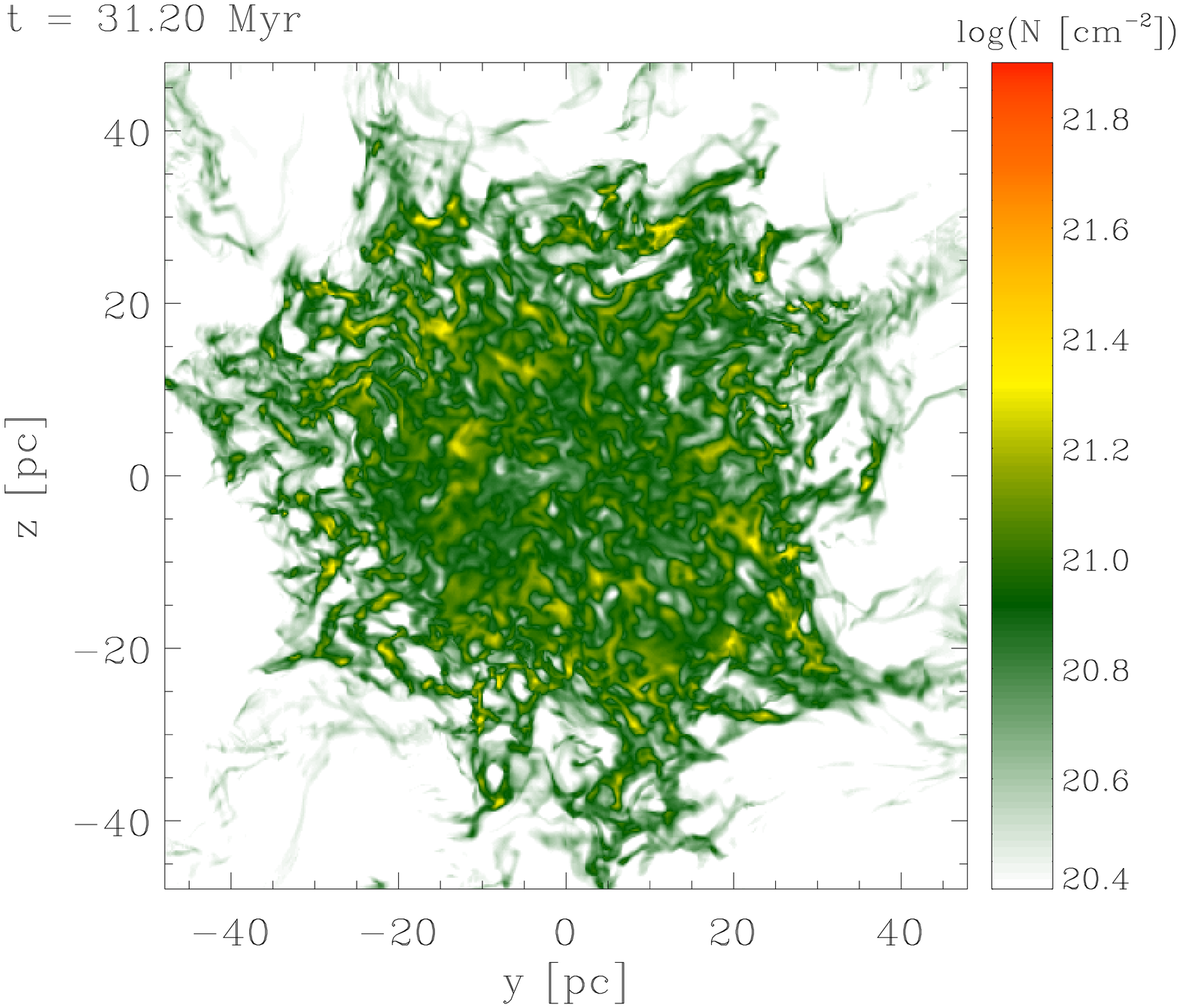}
\includegraphics[width=5.68cm]{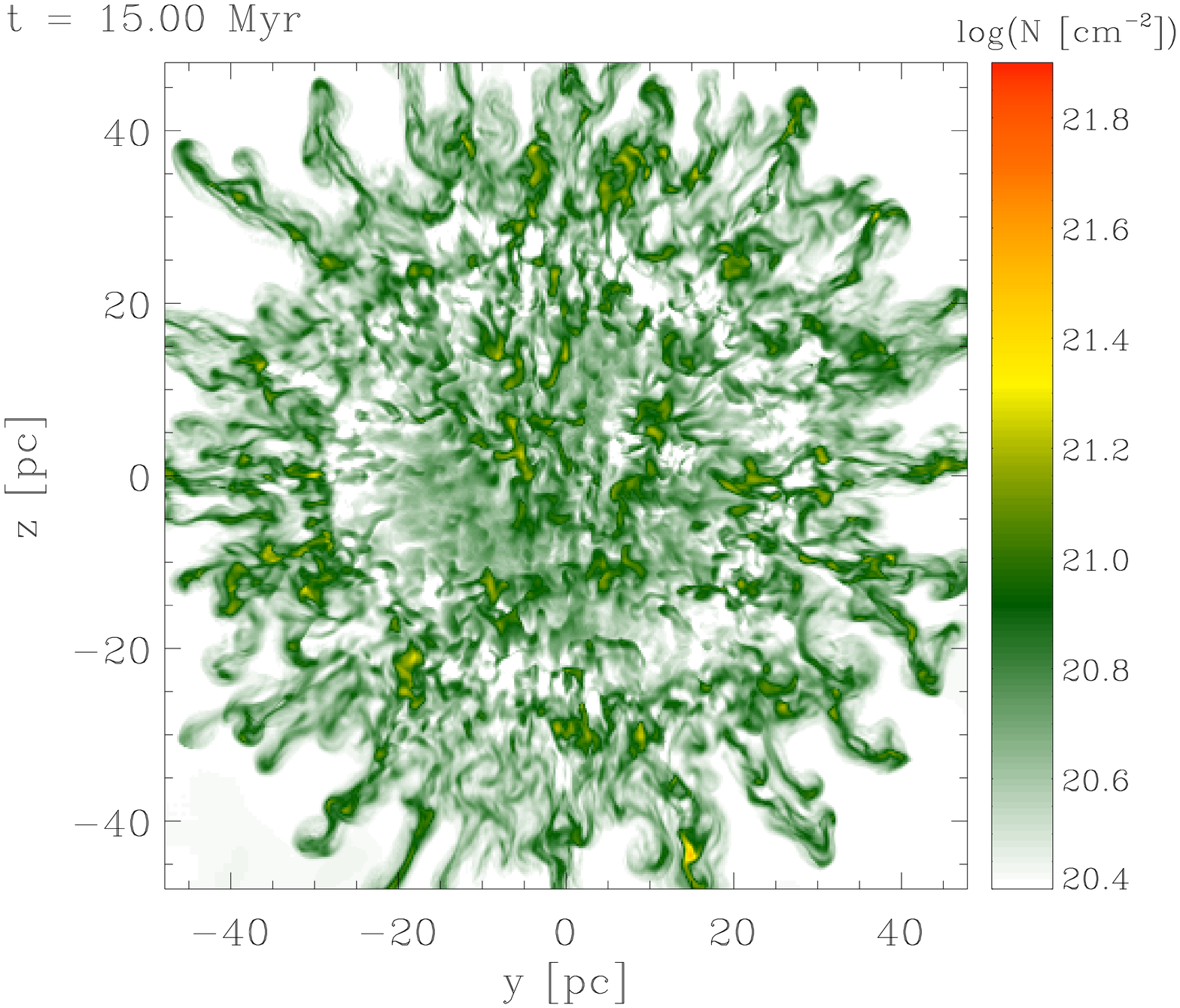}
\includegraphics[width=5.68cm]{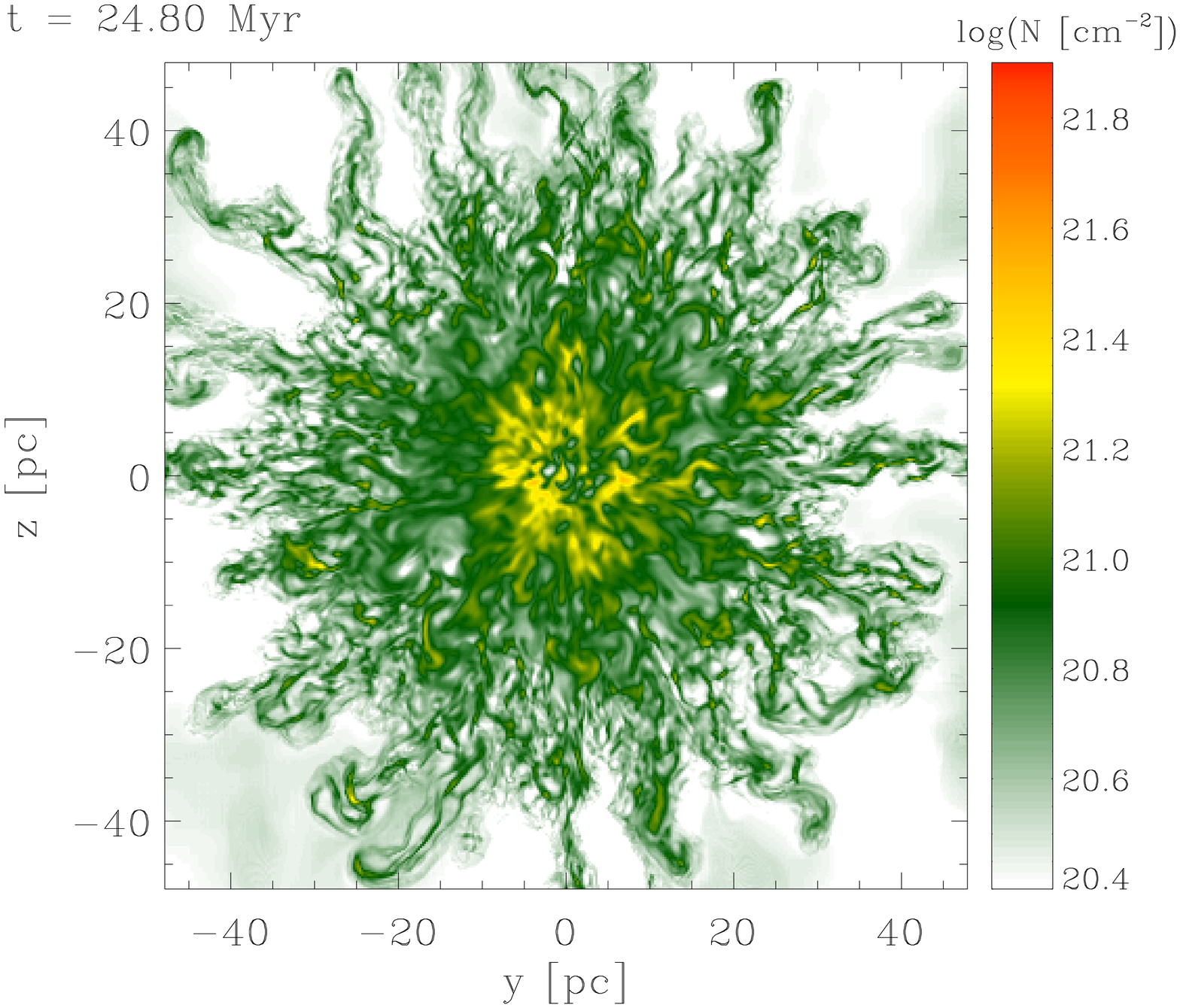}
\includegraphics[width=5.68cm]{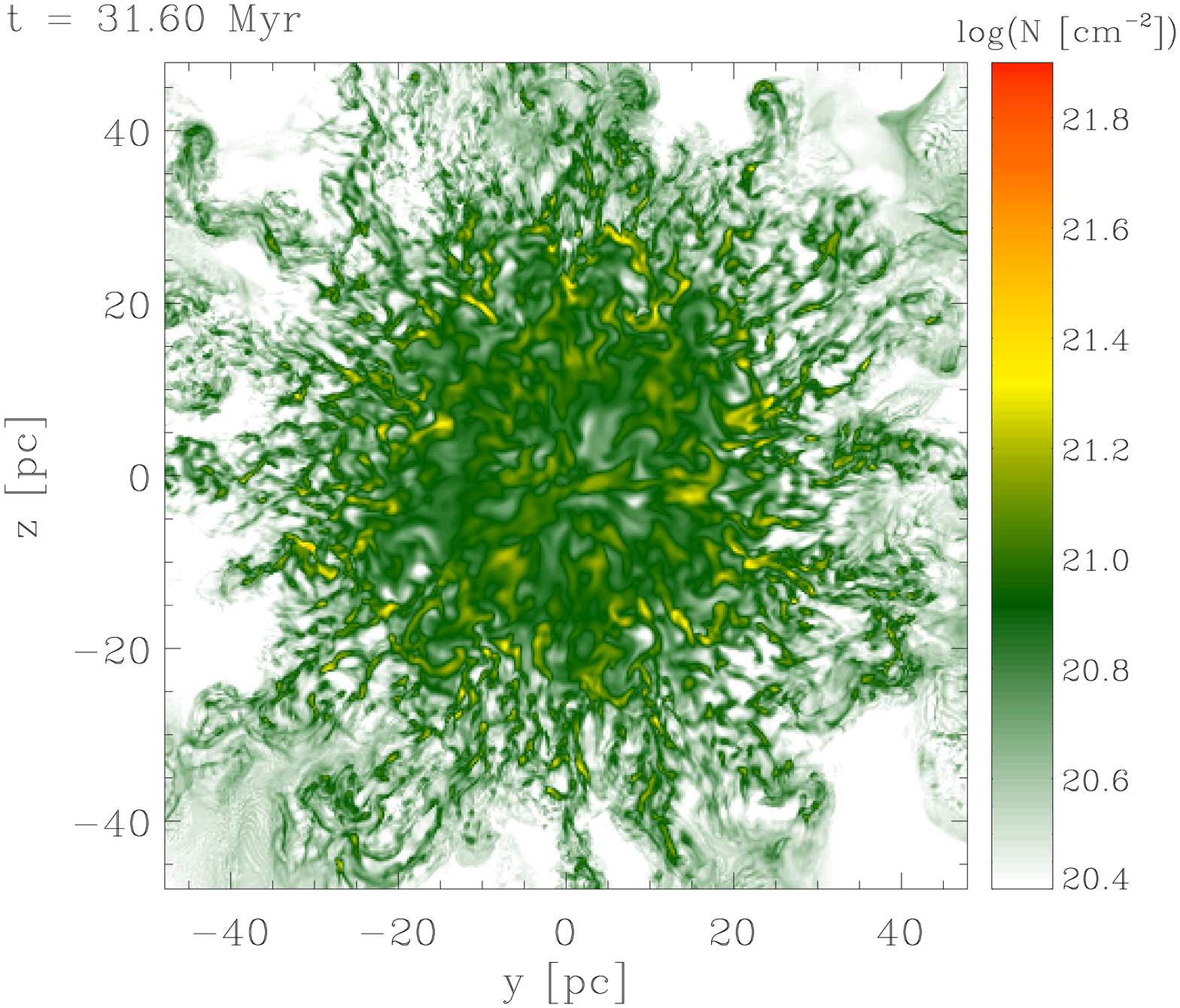}
\caption{Column density of the inner region of the dense cloud viewed face-on at three different output times: $t\sim15$ Myr, $t\sim25$ Myr and $t\sim32$ Myr. The top panels show the results from the run that used the Koyama \& Inutsuka
cooling function, while the bottom panels show the results from the run with the full non-equilibrium treatment. Note that
output times considered here and in Figures~\ref{fig:column_edge}--\ref{fig:vrad} 
below are slightly different in the two runs owing to minor differences in the
timing of the output snapshots produced by \textsc{FLASH} in the two different simulations.}
\label{fig:column_face}
\end{figure*}

\begin{figure*}
\centering
\includegraphics[width=5.68cm]{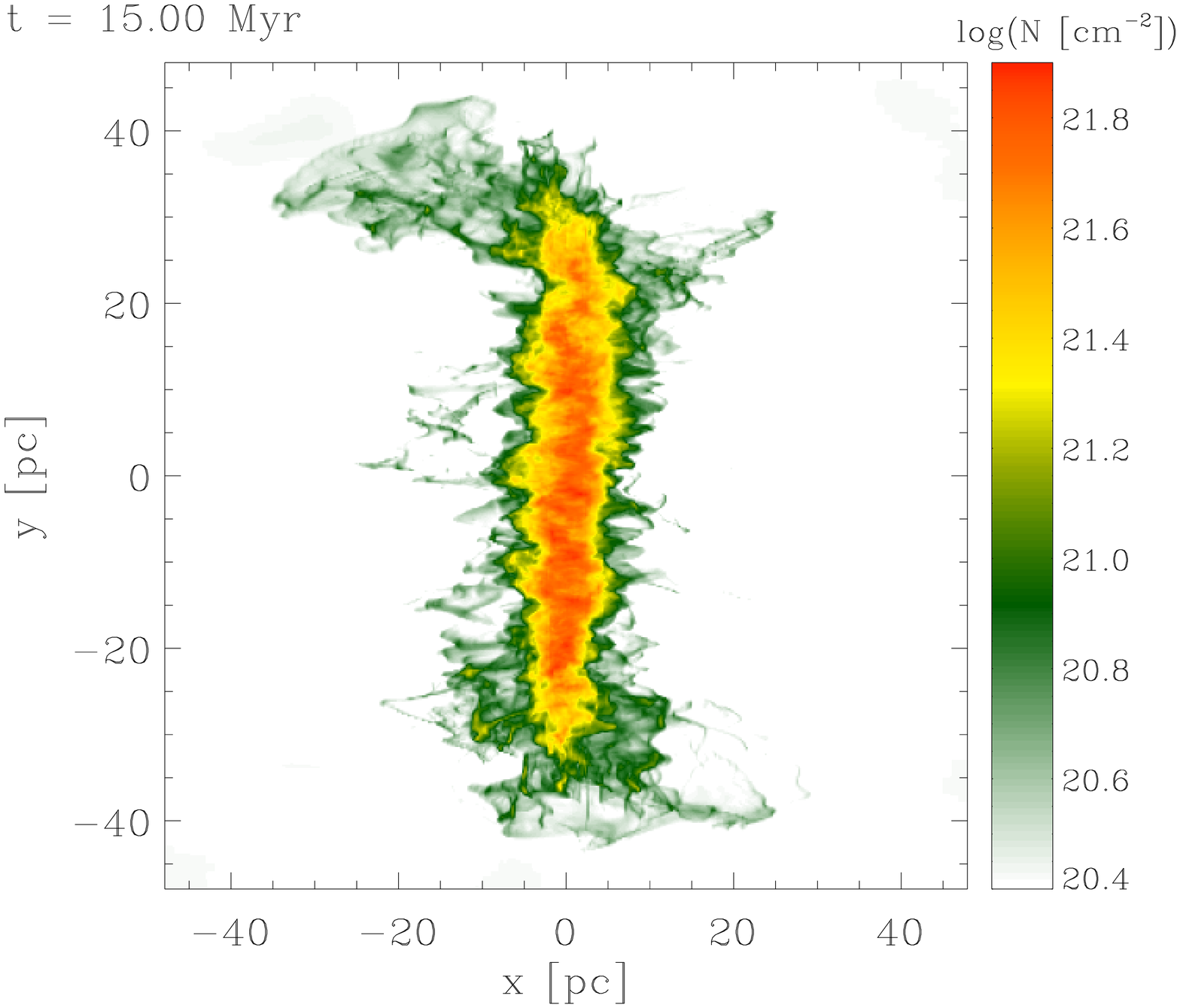}
\includegraphics[width=5.68cm]{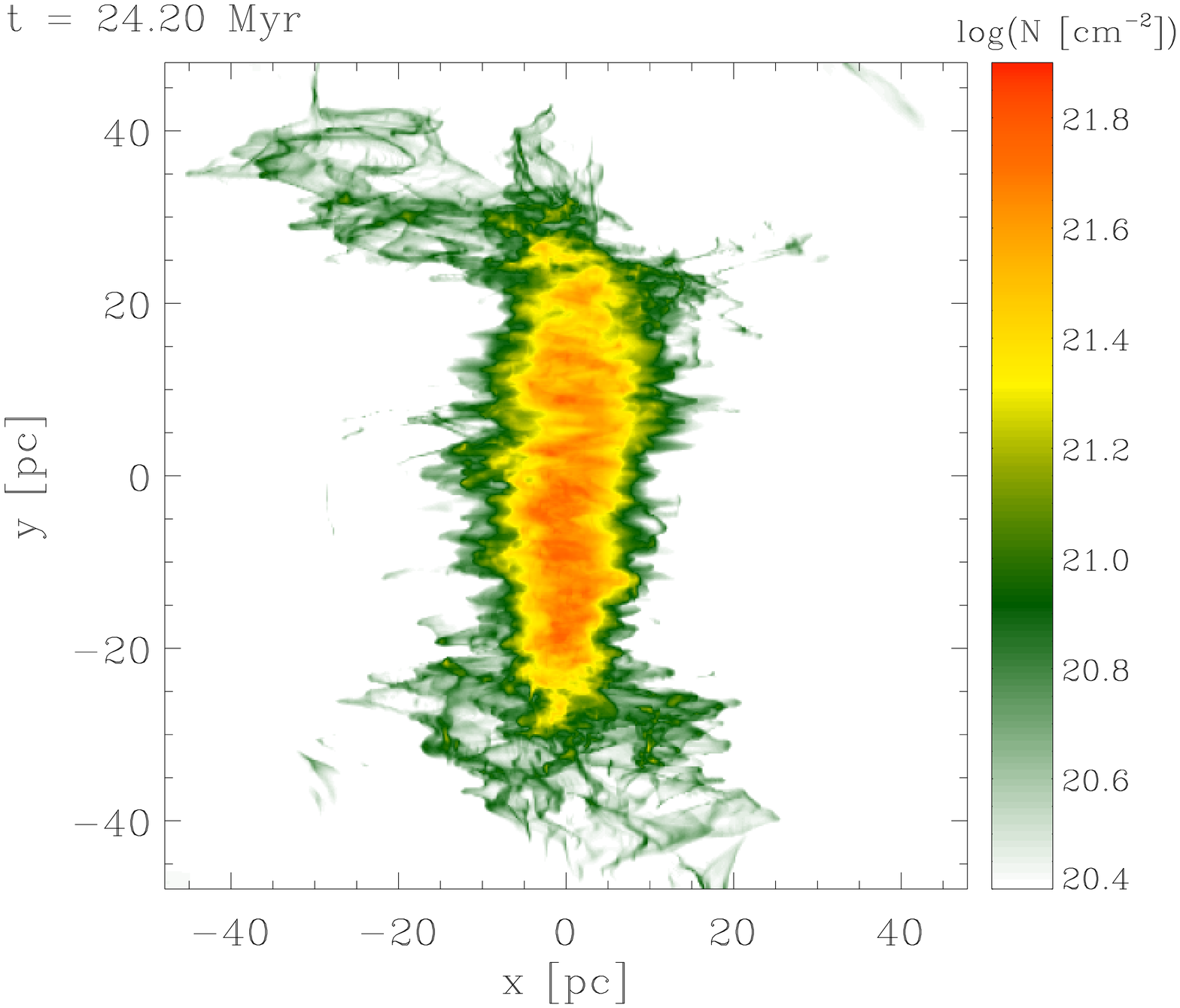}
\includegraphics[width=5.68cm]{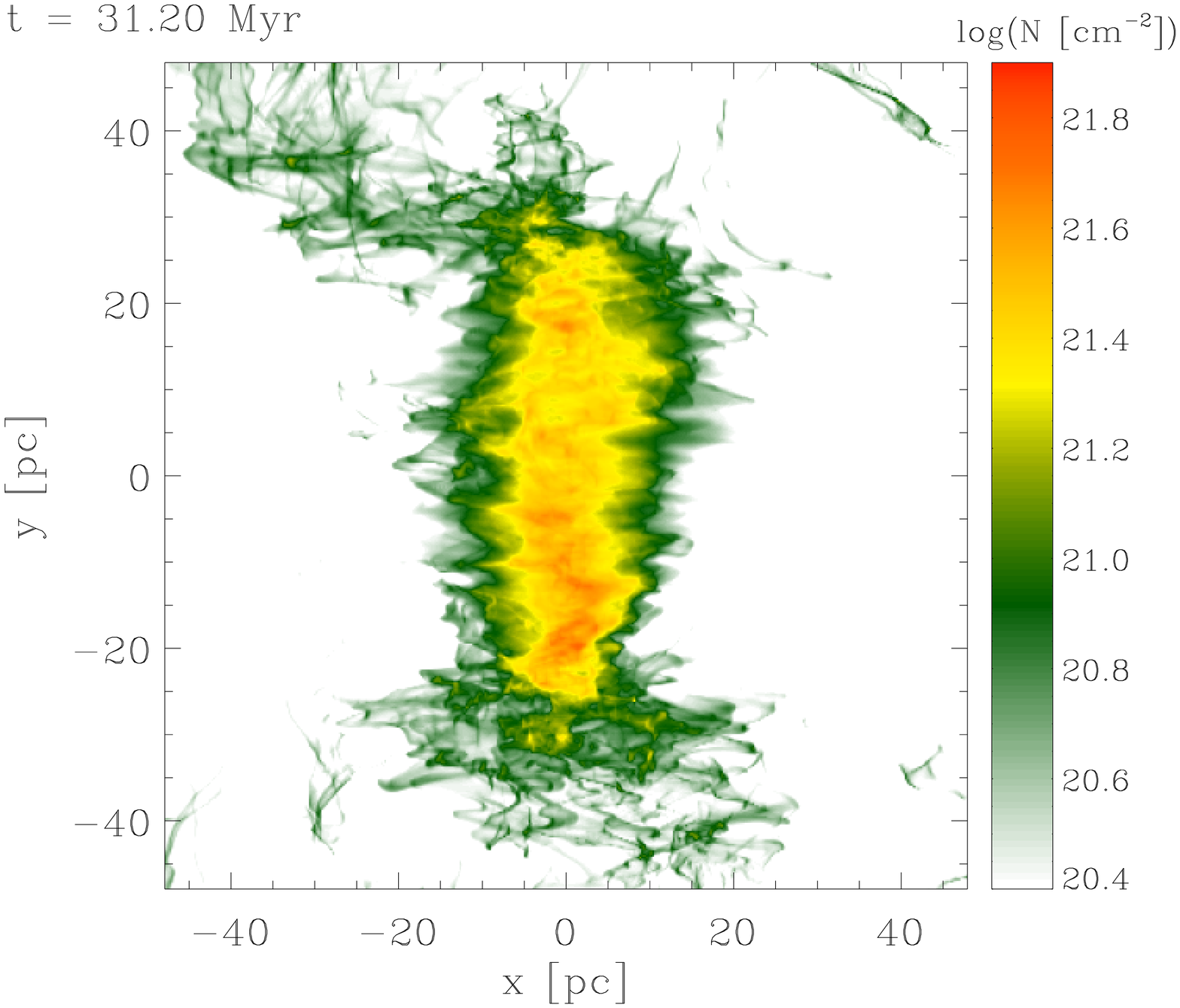}
\includegraphics[width=5.68cm]{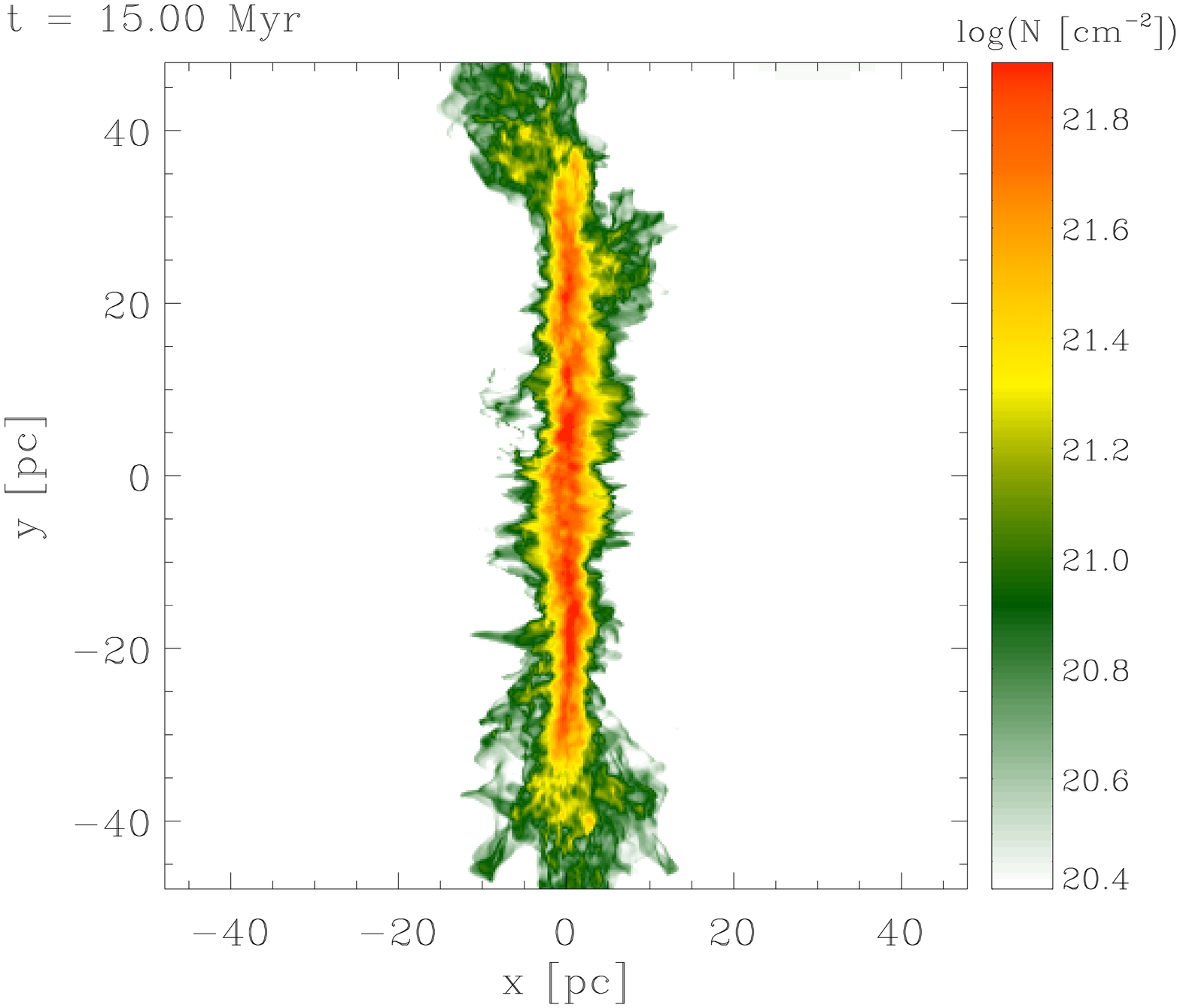}
\includegraphics[width=5.68cm]{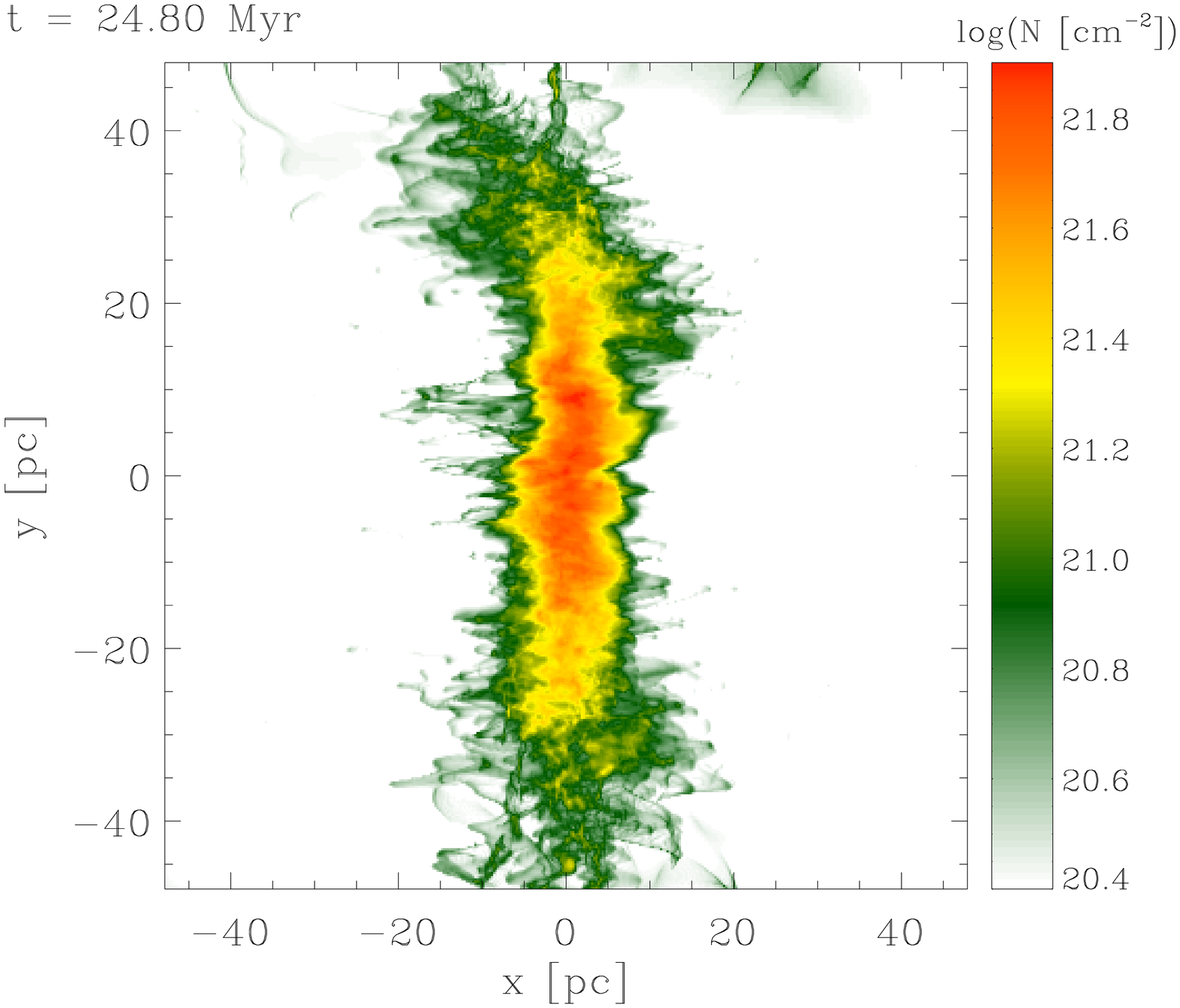}
\includegraphics[width=5.68cm]{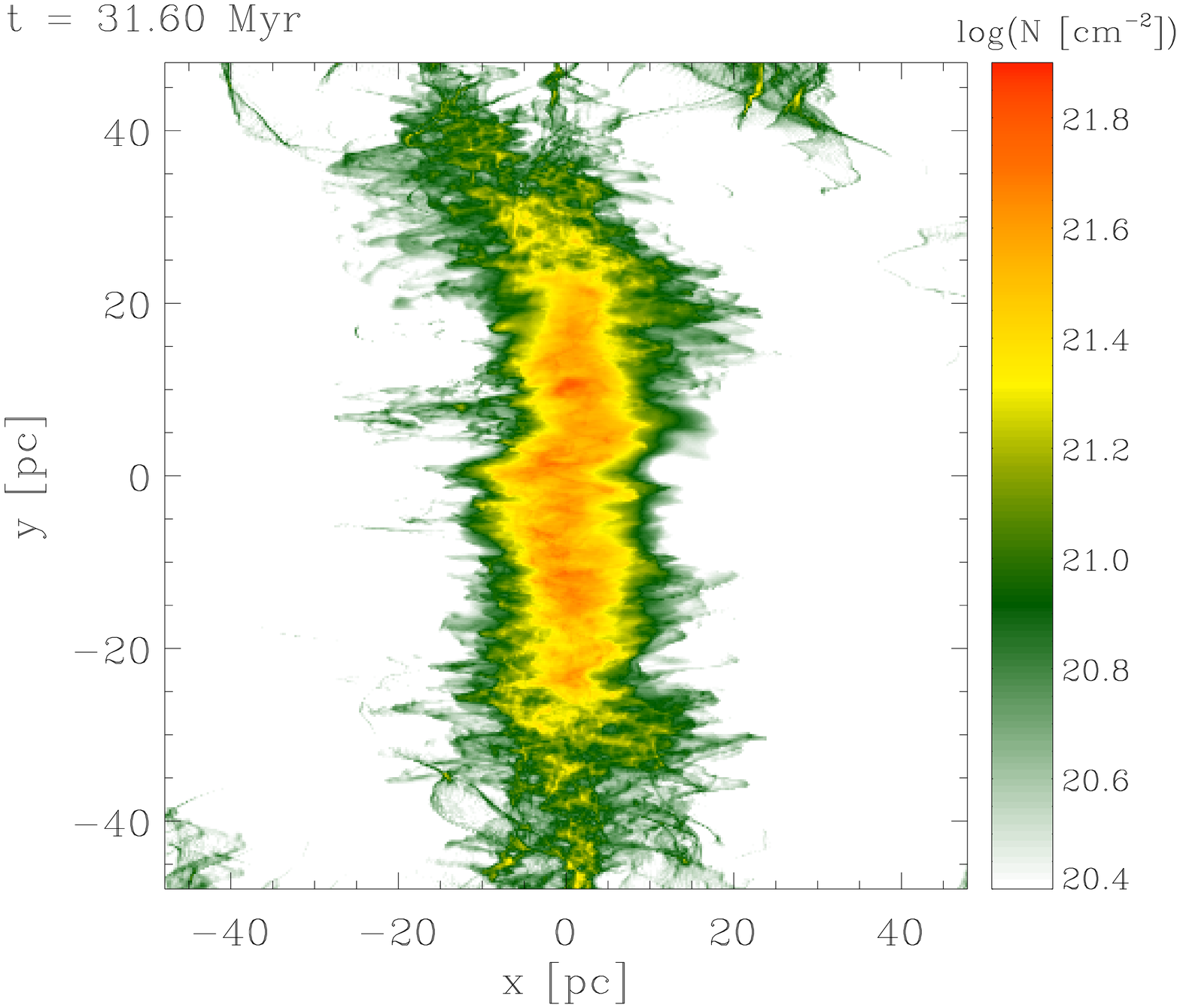}
\caption{Column density of the inner region of the dense cloud viewed edge-on at three different output times: $t\sim15$ Myr, $t\sim25$ Myr and $t\sim32$ Myr. The top panels show the results from the run that used the Koyama \& Inutsuka cooling function, while the bottom panels show the results from the run with the full non-equilibrium treatment.}
\label{fig:column_edge}
\end{figure*}
\subsection{Cloud structure and velocity}

In Figure~\ref{fig:column_face}, we show projections of the column density of hydrogen nuclei, $N$, along the axis of the flow at three comparable output times for both of our simulations. The top panels show the results for the run with the Koyama \& Inutsuka cooling function, while the bottom panels show the results from our non-equilibrium chemistry run. Figure~\ref{fig:column_edge} shows a similar comparison, but for a direction perpendicular to the flow.

The images in Figure~\ref{fig:column_face} show us clearly that the cloud is not a homogeneous entity, but rather is composed of numerous dense clumps embedded in lower density filaments. Moreover, by comparing the results of the two runs, one can see clearly that the cloud morphology is sensitive to the details of the thermal treatment adopted. In the run with the Koyama \& Inutsuka cooling function, the clumps and filaments are contained within a region that is still roughly circular, and that has a radius that is only slightly larger than the initial radius of our inflowing gas ($R \sim 37$~pc at $t = 15 \: {\rm Myr}$, compared with $R = 32$~pc initially). On the other hand, in the non-equilibrium chemistry run, the dense gas occupies a significantly larger region, and is less circular, having an edge that is dominated by long, thin filaments of gas. The larger size of the cloud in the non-equilibrium run is also clearly apparent in the side-on view (Figure~\ref{fig:column_edge}).

\begin{figure*}
\centering
\includegraphics[width=5.68cm]{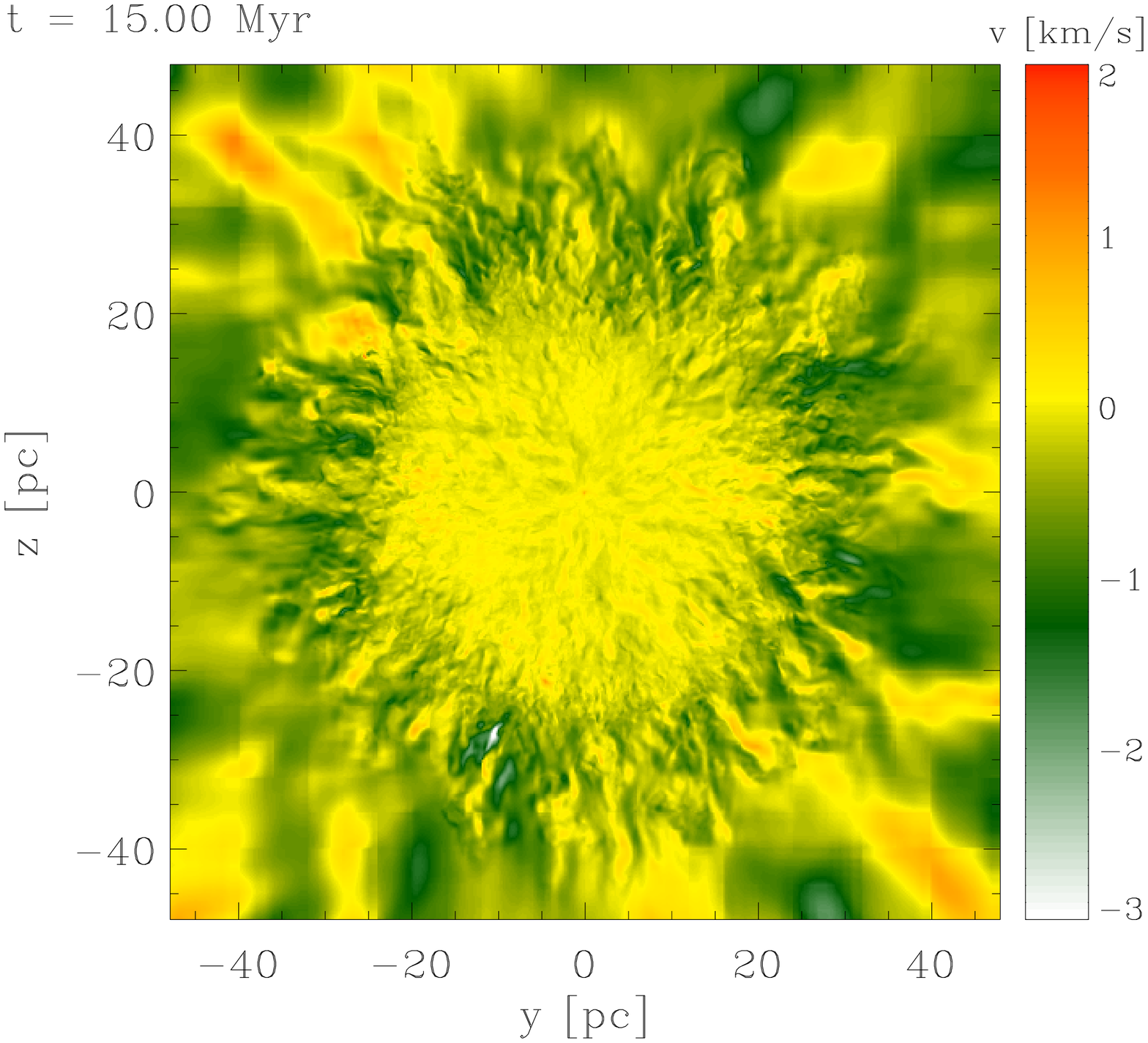} 
\includegraphics[width=5.68cm]{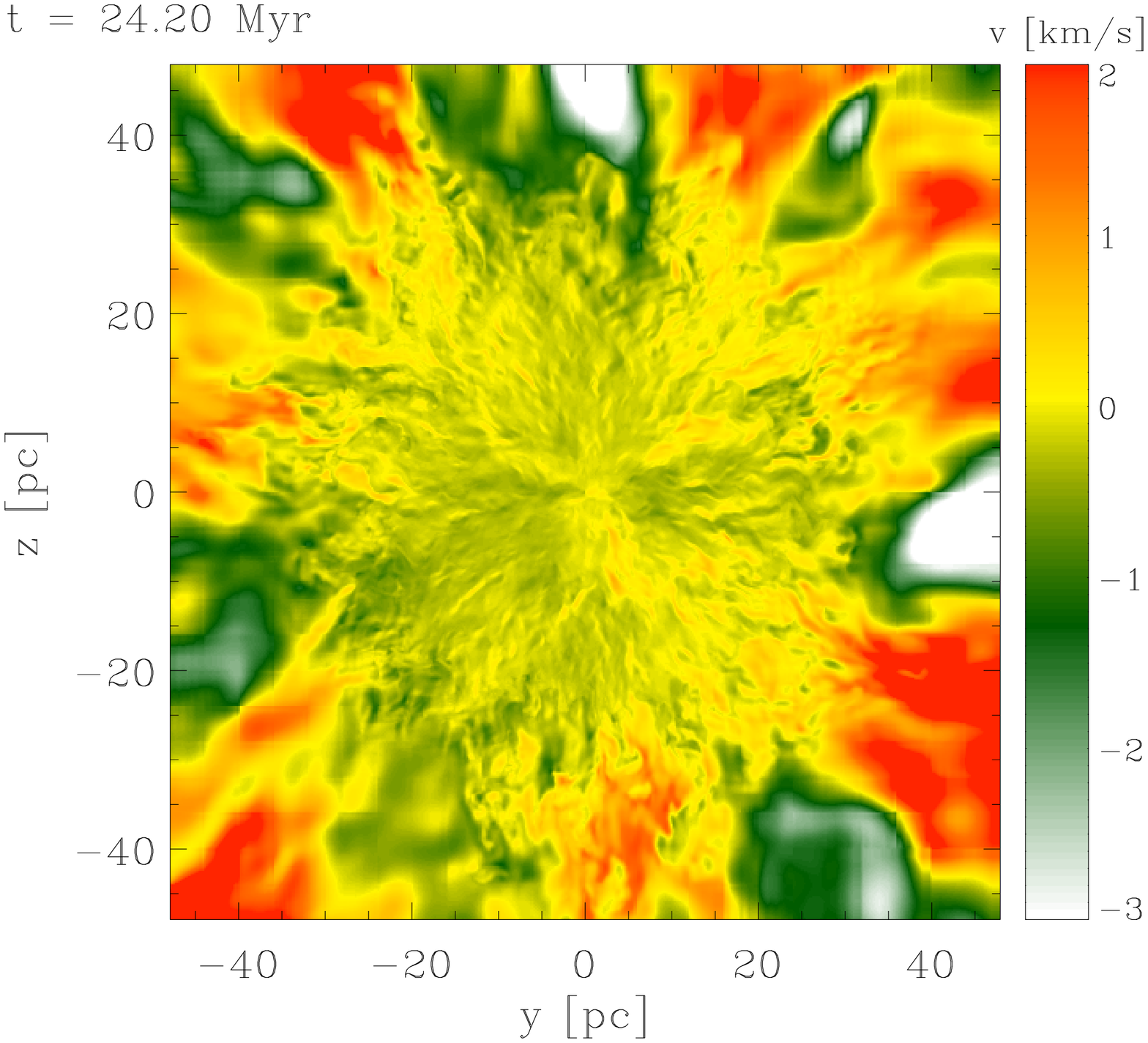}
\includegraphics[width=5.68cm]{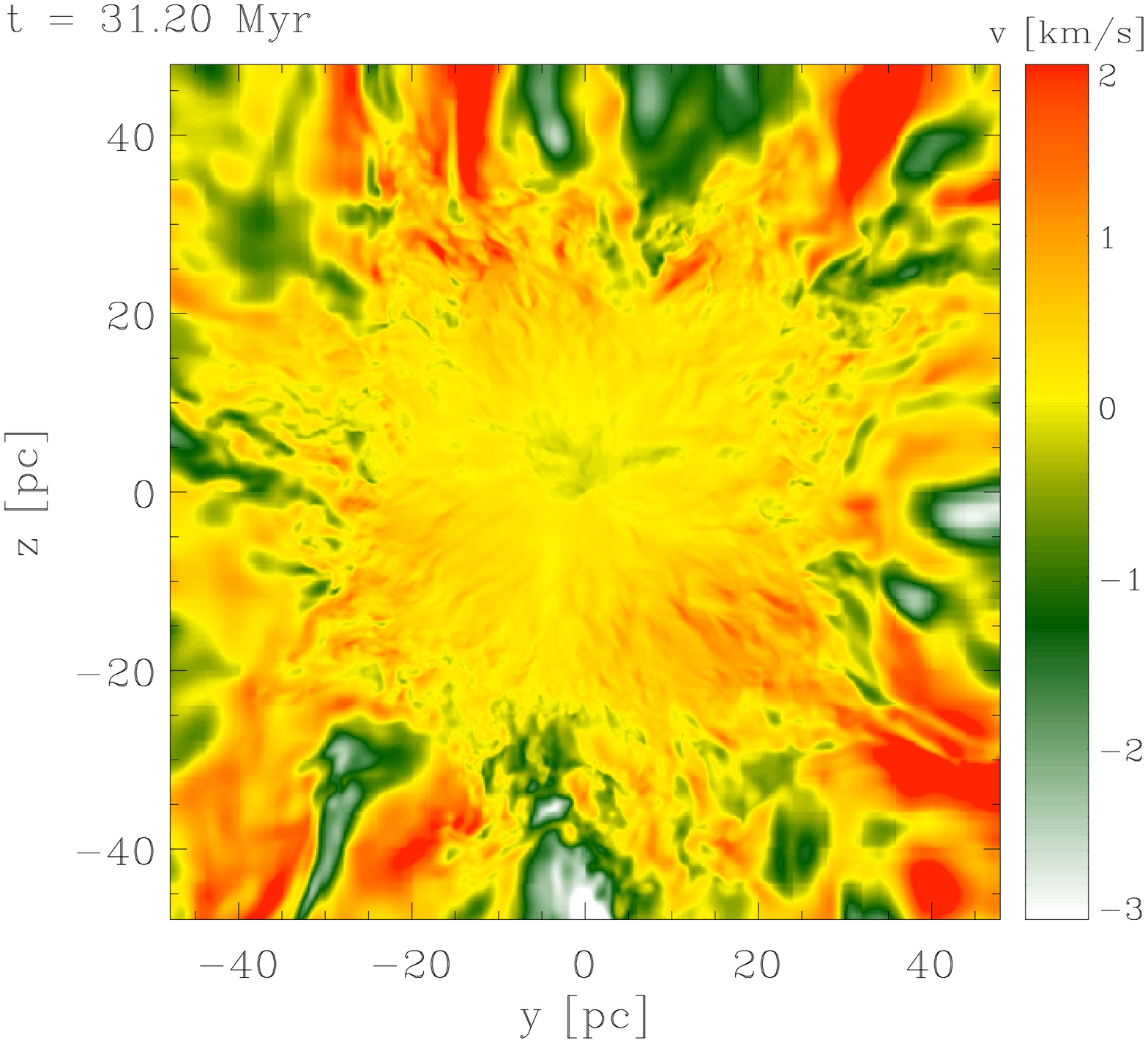}
\includegraphics[width=5.68cm]{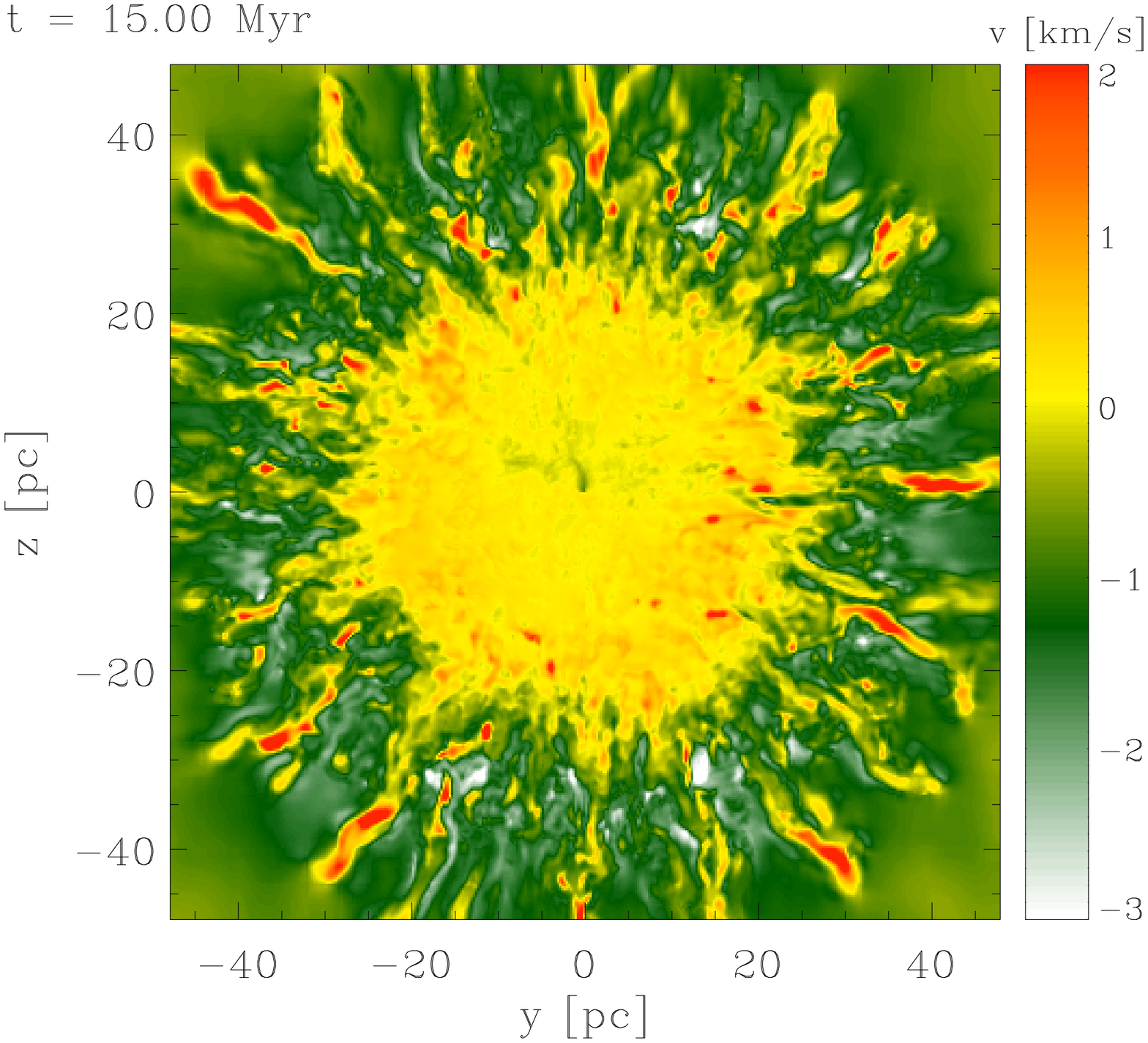}
\includegraphics[width=5.68cm]{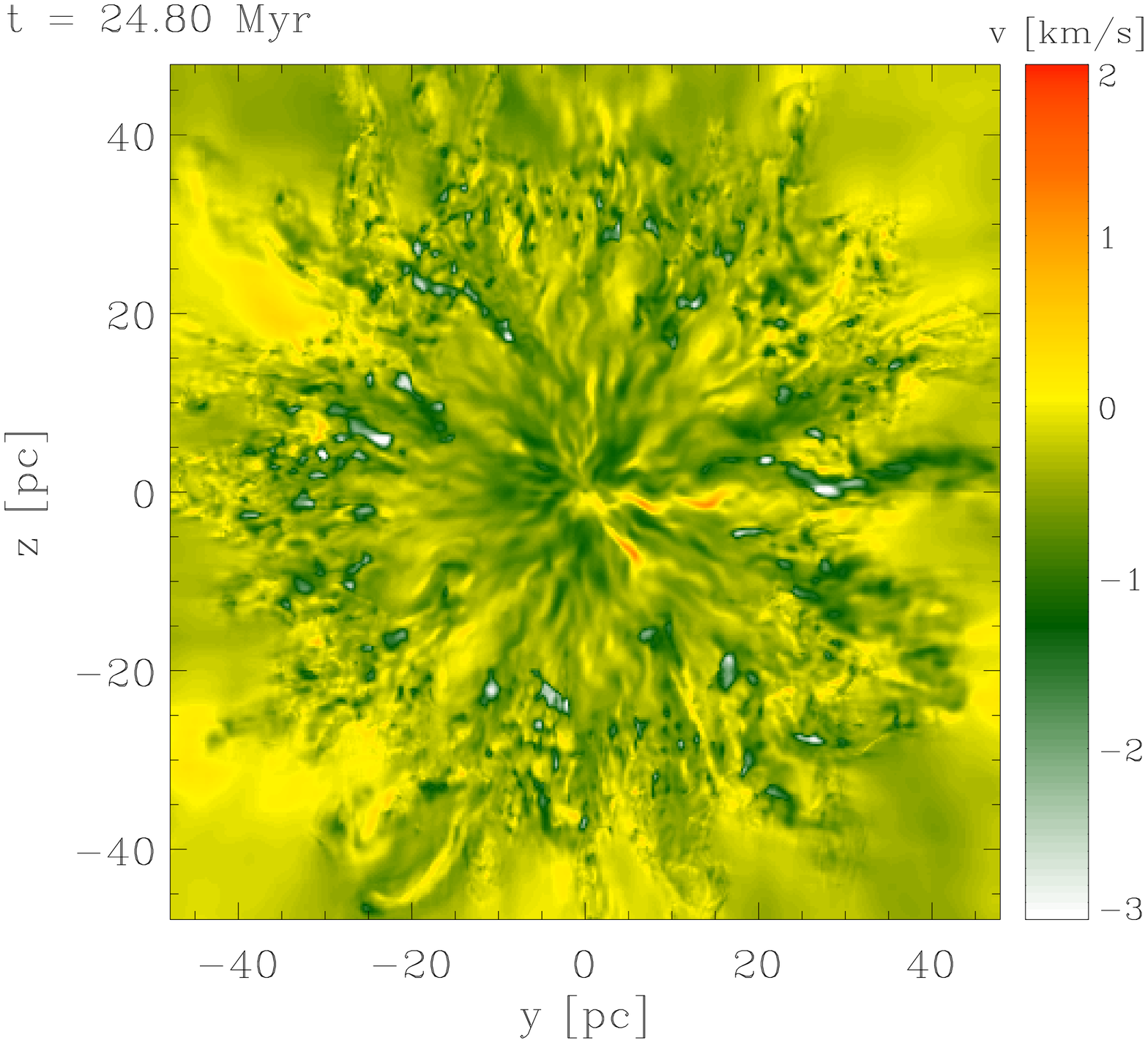}
\includegraphics[width=5.68cm]{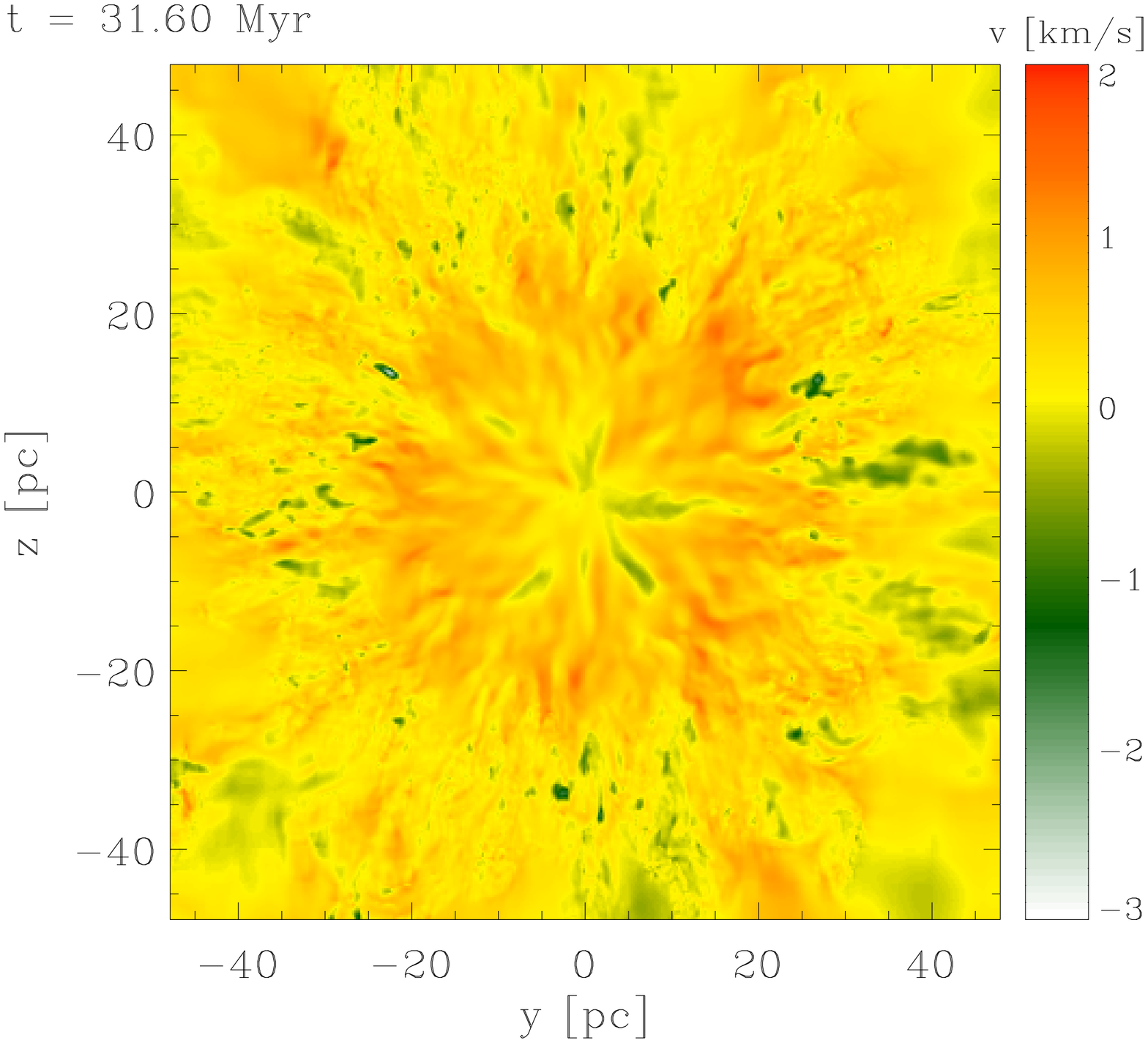}
\caption{Radial velocity of the gas in a slice through the centre of the cloud, relative to the gas at the centre of the cloud. As in Figures~\ref{fig:column_face} and \ref{fig:column_edge}, we plot results from three different output times ($t\sim15$~Myr, $t\sim25$~Myr and $t\sim32$~Myr) for both the simulation using the Koyama \& Inutsuka cooling function (top panels) and the full non-equilibrium treatment (bottom panels).}
\label{fig:vrad}
\end{figure*}

Looking at the velocity distribution of the gas perpendicular to the flow (i.e. the radial velocities shown in Figure~\ref{fig:vrad} for a slice through the centre of the dense cloud), we see that the immediate cause of the difference in morphologies is a difference in the velocity distributions. In the Koyama \& Inutsuka run, the net outward velocity of the dense gas is very small. In the non-equilibrium chemistry run, on the other hand, the gas near the axis of the inflow is relatively static, but the dense gas close to the edges of the distribution is largely flowing outwards. In particular, the gas in the filaments has outward velocities of as much as $2 \: {\rm km} \: {\rm s^{-1}}$.

As the dense gas moves outwards in the non-equilibrium chemistry run, it drags the magnetic field lines along with it. The magnetic tension generated by this disturbance of the field lines exerts an inwards force on the expanding gas, and this force eventually becomes strong enough to reverse the direction of the flow. This effect can be seen quite clearly when we look at the column density distribution and velocity field at $t \sim 25 \: {\rm Myr}$ (the middle panels of Figures~\ref{fig:column_face}--\ref{fig:vrad}). The gas distribution has become more compact, resulting in higher column densities, particularly close to the central axis of the flow, and more of the gas is flowing in than is flowing out. At an even later time ($t \sim 32 \: {\rm Myr}$; right-hand panels in Figures~\ref{fig:column_face}--\ref{fig:vrad}), the flow has ``bounced'' and has begun to re-expand once more. Looking at the results from the run with the Koyama \& Inutsuka cooling function, we see hints of similar behaviour, but in this case both the initial outflow and the subsequent inflow are much weaker.

The root cause of the difference in behaviour between the two runs is the thermal evolution of the shock-heated gas in the central cloud. The two inflowing streams of gas each are moving at a speed of $7 \: {\rm km} \: {\rm s^{-1}}$, and so their relative velocity is $14 \: {\rm km} \: {\rm s^{-1}}$, or around 2.5 times the speed of sound in the warm gas. When the gas collides, it passes through a shock, which heats it and compresses it. We can estimate the post-shock density and temperature by applying the standard shock jump conditions. Since the magnetic field in our simulations is oriented along the flow, it plays no role in determining the post-shock conditions, and the same conditions apply as for a purely hydrodynamical shock. For the density, we therefore have the relationship
\begin{equation}
\frac{\rho_{2}}{\rho_{1}} = \frac{(\gamma + 1) {\cal M}^{2}}{(\gamma - 1) {\cal M}^{2} + 2}, 
\end{equation}
where $\rho_{1}$ is the pre-shock density, $\rho_{2}$ is the post-shock density, and $\gamma$ is the adiabatic index
of the gas. For the temperature, we have the relationship
\begin{equation}
\frac{T_{2}}{T_{1}} = \frac{\left(1 + \frac{\gamma - 1}{2} {\cal M}^{2} \right) \left(\frac{2\gamma}{\gamma - 1} {\cal M}^{2} - 1 
\right)}{{\cal M}^{2} \left(\frac{2\gamma}{\gamma - 1} + \frac{\gamma - 1}{2} \right)},
\end{equation}
where $T_{1}$ and $T_{2}$ are the pre-shock and post-shock temperatures, respectively. For the case of ${\cal M} = 2.5$ and $\gamma = 5/3$, we therefore find that $\rho_{2} \simeq 2.7 \, \rho_{1}$ and $T_{2} \simeq 2.8 \, T_{1}$. The post-shock thermal pressure is therefore a factor of around 7.6 larger than the pre-shock thermal pressure, which itself is the same as the thermal pressure of the surrounding gas not participating in the inflow. The shocked gas is therefore over-pressured relative to its surroundings, and the resulting pressure gradient causes the gas to expand in the directions perpendicular to the inflow. In the inflow direction, it is confined by the ram pressure of the flow. The effectiveness with which this pressure gradient can accelerate gas outwards from the central cloud depends on the length of time the gas remains in this over-pressured state. If we assume that the initial cooling of the gas is isochoric (i.e.\ that there is no change in its density), then it will cease to be over-pressured once its temperature drops below $T_{1} \rho_{1} / \rho_{2}$, which for the case considered above and an initial temperature of 5000~K yields $T \sim 1850$~K.

\begin{figure}
\centering
\includegraphics[width=7.8cm]{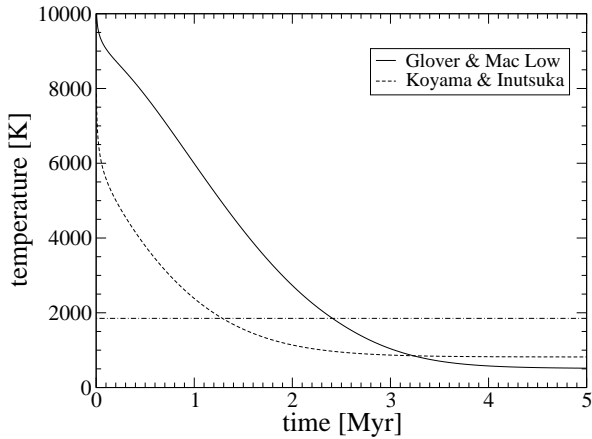} 
\caption{Evolution with time of the temperature of a fluid element with initial density $n_{i} = 2.7 \: {\rm cm^{-3}}$ and
initial temperature $T_{i} = 14000$~K, evolving at constant density. The initial chemical state of the gas, the strength
of the UV radiation field, the cosmic ray ionisation rate and the metallicity are all taken to be the same as in our colliding
flow simulations. The solid line shows the temperature evolution that we obtain when we model the gas using our
non-equilibrium chemical model and cooling function, while the dashed line shows the results that we obtain when 
we use the Koyama \& Inutsuka cooling function. The horizontal dash-dotted line shows the temperature at which the thermal
pressure of the gas is the same as that of the unperturbed WNM in our colliding flow simulations.}
\label{fig:tevol}
\end{figure}

In Figure~\ref{fig:tevol}, we show how the temperature of a parcel of gas with density $n = 2.7 \: {\rm cm^{-3}}$ and temperature $T = 14000$~K changes as a function of time when we model heating and cooling using the Koyama \& Inutsuka cooling function (dashed line) or our full non-equilibrium treatment (solid line). In the latter case, we take the initial chemical state of the gas to be the same as in our simulations. We see that when we use the Koyama \& Inutsuka cooling function, the gas cools very rapidly at early times, returning to its original temperature after only 0.3~Myr. Subsequently, it cools more slowly, and it ceases to be over-pressured with respect to the unperturbed gas after around 1.3~Myr. On the other hand, when we model the chemistry and cooling of the gas using our non-equilibrium model, we see that it takes considerably longer for the gas too cool. In this case, the temperature of the gas returns to its original value after around 1.2~Myr, and the gas remains over-pressured until $t \sim 2.45$~Myr. These results demonstrate that when we use our non-equilibrium treatment to model the cooling of the gas, it remains significantly over-pressured for a much longer period than when we use the Koyama \& Inutsuka cooling function. In the former case, therefore, the gas is accelerated by an outward-pointing pressure gradient for a longer period of time, and hence attains a significantly higher outward velocity, accounting for the differences we see in the morphology of the cloud and the velocity structure of the gas. 

\subsection{Clump properties}
\label{clump}

Finally, we explore whether the differences in cloud morphology and in the velocity field of the gas lead to significant differences in the statistical properties of the dense clumps formed in the flow. Although the fact that we do not include self-gravity in our models prevents us from following the further evolution of this dense gas in detail, we know from previous studies that it is the ongoing growth and merger of these dense clumps that eventually leads to the formation of gravitationally unstable pre-stellar cores and, ultimately, stars \citep[see e.g.][]{robi09,vaz11}. Differences in the clump properties at an early stage may therefore be indicative of differences in the ability of the clouds to form stars.

\begin{figure*}
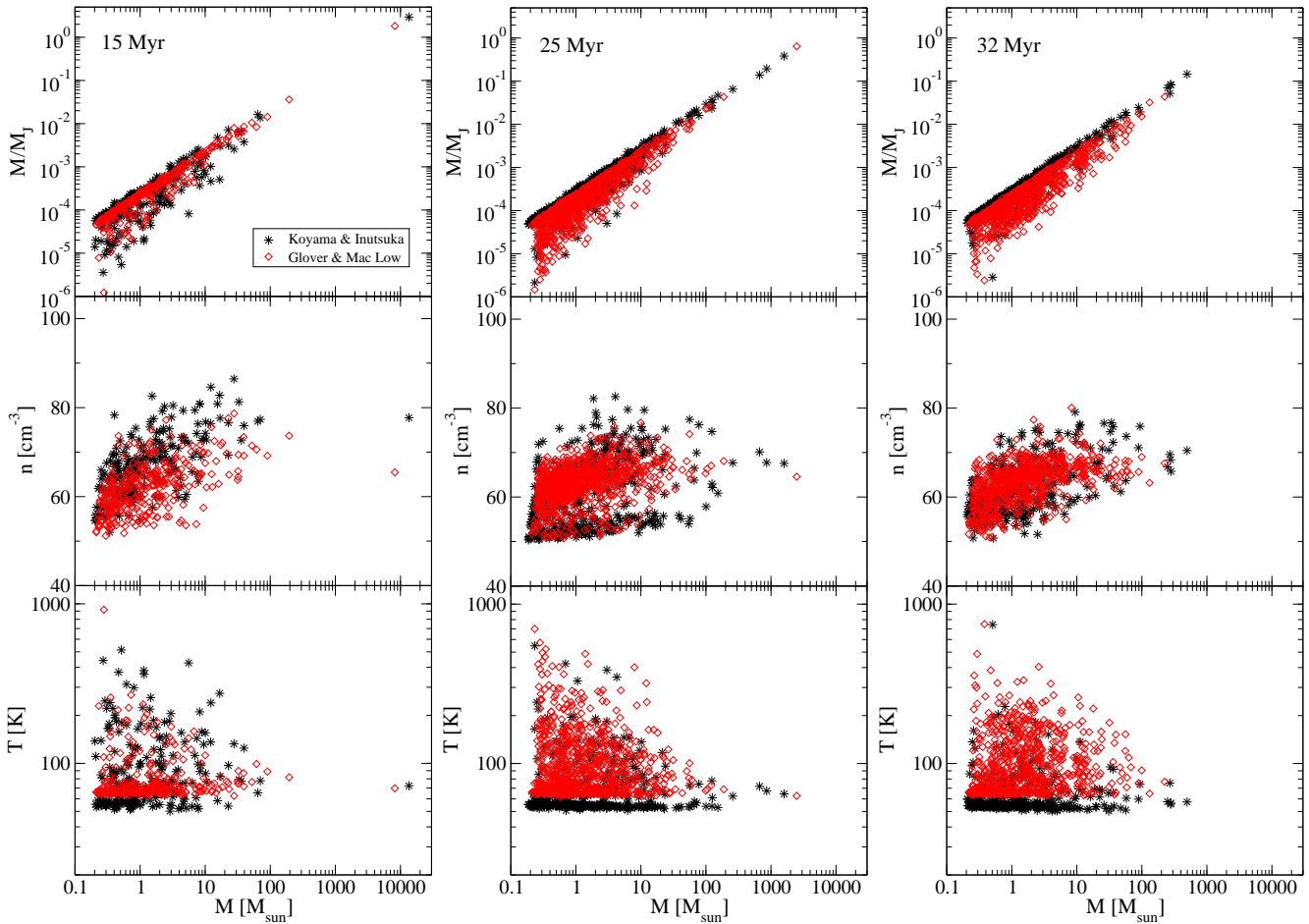

\centering
\includegraphics[width=5.8cm]{f08a.eps}
\includegraphics[width=5.8cm]{f08b.eps}
\includegraphics[width=5.8cm]{f08c.eps}
\caption{Properties of the set of dense clumps identified as described in Section~\ref{clump}. Results are shown for three output times: $t\sim15$~Myr (left-hand panels), $t\sim25$~Myr (central panels) and $t\sim32$~Myr (right-hand panels). The upper row of panels shows the ratio of the clump mass to the local Jeans mass, $M_{\rm J}$, the central row shows the mean density of the clumps and the lower row shows the mean temperature. In each case, we plot these quantities as a function of the total clump mass. The red symbols indicate the results that we obtain in the run with our non-equilibrium chemical model, while the black symbols indicate the results that we obtain when we use the Koyama \& Inutsuka cooling function. }

\label{fig:stat}
\end{figure*}

We identify clumps in our simulations by searching for connected regions with densities above 50 cm$^{-3}$ using the SEARCH3D routine of IDL. This routine is consecutively applied on the peak densities in the simulation excluding the already extracted clumps until we identified all connected regions with densities above 50 cm$^{-3}$. For reasons of computational efficiency, we restrict our search for clumps to radial distances $R \leq 40$~pc from the central axis of the flow, but from Figure~\ref{fig:column_face} we can clearly see that this region contains almost the entire mass of dense gas, and so we are unlikely to miss many clumps. In Figure~\ref{fig:stat}, we show some of the averaged internal properties of the clumps at times $t\sim15$ Myr, $t\sim25$ Myr, and $t\sim32$ Myr for both models. 

The top panels of these figures show the ratio of the clump mass $M$ to the local Jeans mass $M_{\rm J}$, plotted as a function of the clump mass. The masses of the clumps span a wide range, 0.1--10$^3~\rm{M}_{\sun}$. The fact that we find no clumps smaller than around 0.1 M$_\odot$ is a consequence of the limited numerical resolution of our simulations, which suppresses the formation of extremely small scale structures. However, we see that even with our current resolution, the vast majority of the clumps are not self-gravitating, since they have $M / M_{\rm J} < 1$. The only exceptions are a couple of clumps present at $t = 15$~Myr that have $M \sim 1$--2~$M_{\rm J}$. We therefore see that in order to form a significant number of self-gravitating clumps -- a necessary pre-requisite for star formation -- the compressions produced by the collision of the flows and the consequent thermal instability are not sufficient; some form of large-scale collapse of the cloud is also required. This is also evident in simulations including self-gravity, where in the case of a critical setup the presence of magnetic fields suppress star formation \citep[]{vaz11}. Similar results have been found in a number of other studies \citep[see e.g.][]{ki02,heit05,vaz07}.

Comparing the clump properties in the different models, we see that the mean density of the clumps is not particularly sensitive to the way in which the cooling of the gas is modelled, although there is a tentative hint that the clumps in the simulation with the Koyama \& Inutsuka cooling function may be slightly denser than in those in the other simulation. A more pronounced difference is apparent in the mean temperature of the clumps, which we find is roughly 10~K higher when we use the non-equilibrium treatment than when we use the Koyama \& Inutsuka cooling function. This difference is due to the difference in the photoelectric heating rate assumed in the two simulations, as noted in Section~\ref{dense} above, and leads to minor differences in the Jeans masses that we determine for our clumps, as one can see from the upper panels in Figure~\ref{fig:stat}. However, it should be noted that {\em both} simulations likely overestimate the temperature of these dense clumps, owing to their neglect of the effects of dust shielding \citep[c.f.][]{gc12,gc12b,clark12}.

\section{Summary}

We have presented the results of a study that examines the influence of two different thermal models on the formation of cold, dense clouds within converging flows of warm atomic gas, and on the nature of the clumps that form within these clouds. To do this, we performed high-resolution 3D MHD simulations using the massively parallel code \textsc{FLASH}, modified to include a detailed treatment of atomic and molecular cooling, and a simplified but accurate treatment of the most important hydrogen chemistry \citep[]{mm12}. We directly compare the results obtained from this model with those that we obtain if we use a simplified cooling function, taken from the work of \citet{ki02}, that has been used in a number of other studies of cloud formation in converging flows \citep[see e.g.][]{robi09}.

We find that the density and temperature PDFs produced in the two simulations are qualitatively similar, although some minor quantitative differences exist. In common with previous work \citep[e.g.][]{vaz07,hen08,robi09} we find that in addition to clear CNM and WNM phases, there is also a significant fraction of mass located in the thermally unstable region between these two phases. The temperatures recovered for the CNM and WNM phases are slightly smaller in the simulation run with the Koyama \& Inutsuka cooling function, as their approach significantly overestimates the cooling rate of hot ($T\sim 10^{4} \: {\rm K}$) gas and underestimates the photoelectric heating rate in the CNM regime by a factor of 2.5. In the Koyama \& Inutsuka model, most of the gas is in thermal equilibrium, while in the full non-equilibrium model, significant deviations from thermal equilibrium are apparent for gas in the density range $1<n<30 \: {\rm cm^{-3}}$. In this intermediate density regime, the photoelectric heating efficiency has not yet reached its limiting value and increases with increasing electron number density. Due to the relatively long recombination timescale, the gas is slightly more ionised than it would be in chemical equilibrium, and hence is heated slightly more efficiently.

We have also shown that the cloud morphology is sensitive to the choice of the thermal treatment. The cloud formed in the non-equilibrium chemistry run is larger and more filamentary than the cloud formed in the run with the Koyama \& Inutsuka cooling function. This difference in morphology is caused by a difference in the velocity distribution of the gas, which itself can be understood as a consequence of the difference in the cooling time of hot gas in the two models. In the Koyama \& Inutsuka model, the high temperature, shock-heated gas cools very rapidly, and the gas remains over-pressure with respect to its surroundings for only a very short time. With the full non-equilibrium treatment, however, the cooling time is significantly longer and hence the cloud remains over-pressured for longer. Consequently, the gas is accelerated outwards more efficiently, resulting in a larger, more disordered gas distribution.

Finally, we have investigated whether the properties of the clumps that form in the clouds differ significantly between the two runs. We have identified clumps with masses in a range of 0.1--10$^3~\rm{M}_{\sun}$ in both simulations, but find that almost all of these structures are not self-gravitating, having $M / M_{\rm J} < 1$. This suggests that the compressions produced by the collision of the flows followed by thermal instability are not sufficient to lead to the formation of gravitationally unstable pre-stellar cores and stars. 
In order for star formation to take place, some form of large-scale collapse or compression of the cloud in a direction perpendicular to the inflow appears to be required. One possible source for this compression is the global gravitational contraction of the shocked cloud, as previously noted in \citet{ki02}, \citet{heit05}, and \citet{vaz07}. Another possibility is  compression due to the effects of orbit crossing in a spiral arm shock, as discussed in \citet{bonnell13}. Most of the properties of the clumps are very similar in our two models, with the most significant difference being a systematic offset of roughly 10~K between the mean clump temperatures in the Koyama \& Inutsuka model and those in the non-equilibrium model. However, it should be noted that both models overestimate the temperature of the clumps, owing to their neglect of the effects of dust shielding.

\section*{Acknowledgments}

The authors would like to thank the referee, Ian Bonnell, for valuable comments that have helped us to improve the manuscript. M.M.\ acknowledges financial support by the International Max Planck Research School for Astronomy and Cosmic Physics at the University of Heidelberg (IMPRS-HD) and the Heidelberg Graduate School of Fundamental Physics (HGSFP). The HGSFP is funded by the Excellence Initiative of the German Research Foundation DFG GSC 129/ 1. M.M.\ further acknowledges support from the Ministry of Education,
Science and Technological Development of the Republic of Serbia through the project No. 176021, ``Visible and Invisible Matter in Nearby Galaxies: Theory and Observations" and support provided by the European Commission through FP7 project BELISSIMA (BELgrade Initiative for Space Science,
Instrumentation and Modelling in Astrophysics, call FP7-REGPOT-2010-5,
contract no. 256772). S.C.O.G.\ and R.S.K.\ acknowledge financial support from the {\em Baden-W{\"u}rttemberg Stiftung} via the contract research (grant P-LS-SPII-18) in the program ``Internationale Spitzenforschung". R.S.K. furthermore acknowledges support from the {\em Deutsche Forschungsgemeinschaft} as part of the collaborative research project SFB 881 ``The Milky Way System" (sub-projects B1, B2, B3, and B4) and the priority program SPP 1573 ``Physics of the ISM" (grant KL 1358/14-1). R.B. acknowledges funding by DFG via the grants BA 3706/1-1 and BA 3706/3-1 and funding by the DAAD through the PROALMEX programme (project ID 51301207). The simulations used computational resources from the HLRBII project pr32hu at Leibniz Rechenzentrum Garching. The software used in this work was in part developed by the DOE-supported ASC/Alliance Center for Astrophysical Thermonuclear Flashes at the University of Chicago.

\bsp

\label{lastpage}


\begin{thebibliography}{99}

\bibitem[Abel et al.(1997)]{a97}
Abel, T., Anninos, P., Zhang, Y., Norman, M.~L., 1997, New Astron., 2, 181
\bibitem[\protect\citeauthoryear{Arzoumanian et al.}{2011}]{arz11}
Arzoumanian, D., Andr\'{e}, Ph., Didelon, P., et al., 2011, A\&A, 529L, 6A
\bibitem[\protect\citeauthoryear{Audit \& Hennebelle}{2005}]{audit05}
Audit, E., Hennebelle, P., 2005, A\&A, 433, 1
\bibitem[Bakes \& Tielens(1994)]{bt94}
Bakes, E.~L.~O., \& Tielens, A.~G.~G.~M.\ 1994, ApJ, 427, 822
\bibitem[\protect\citeauthoryear{Ballesteros-Paredes et al.}{1999a}]{bp99a}
Ballesteros-Paredes J., Hartmann L., V\'{a}zquez-Semadeni E., 1999a, ApJ, 527, 285
\bibitem[\protect\citeauthoryear{Ballesteros-Paredes et al.}{1999b}]{bp99b}
Ballesteros-Paredes J., V\'{a}zquez-Semadeni E., Scalo, J., 1999b, ApJ, 515, 286
\bibitem[\protect\citeauthoryear{Banerjee et al.}{2009}]{robi09}
Banerjee, R., V\'{a}zquez-Semadeni, E., Hennebelle, P., Klessen, R. S., 2009, MNRAS, 398, 1082
\bibitem[Beck(2001)]{beck01}
Beck, R., 2001, Space Sci.\ Rev.\ 99, 243
\bibitem[\protect\citeauthoryear{Bonnell, Dobbs \& Smith}{2013}]{bonnell13}
Bonnell, I.~A., Dobbs, C.~L., Smith, R.~J., 2013, MNRAS, in press; arXiv:1301.1041
\bibitem[\protect\citeauthoryear{Bouchut, Klingenberg \& Waagan}{2007}]{bkw07}
Bouchut, F., Klingenberg, C., Waagan, K., 2007, Numerische Mathematik, 108, 7
\bibitem[\protect\citeauthoryear{Bouchut, Klingenberg \& Waagan}{2010}]{bkw10}
Bouchut, F., Klingenberg, C., Waagan, K., 2010, Numerische Mathematik, 115, 647
\bibitem[\protect\citeauthoryear{Brown et al.}{1989}]{dvode}
Brown P. N., Byrne G. D., Hindmarsh A. C., 1989, SIAM J. Sci. Stat. Comput., 10, 1038
\bibitem[Cen(1992)]{cen92}
Cen, R., 1992, ApJS, 78, 341
\bibitem[Clark et~al.(2012)]{clark12}
Clark, P.~C., Glover, S.~C.~O., Klessen, R.~S., \& Bonnell, I.~A., 2012, MNRAS, 424, 2599
\bibitem[\protect\citeauthoryear{Elmegreen}{2000}]{el00}
Elmegreen B. G., 2000, ApJ, 530, 277
\bibitem[\protect\citeauthoryear{Elmegreen \& Scalo}{2004}]{es04}
Elmegreen B. G., Scalo J., 2004, ARA\&A, 42, 211
\bibitem[\protect\citeauthoryear{Falgarone et al.}{1992}]{fal92}
Falgarone, E., Puget, J. L., Perault, M., 1992, A\&A, 257, 715
\bibitem[\protect\citeauthoryear{Federrath et al.}{2011}]{fed11}
Federrath, C., Sur, S., Schleicher, D. R. G., Banerjee, R., Klessen, R. S., 2011, ApJ, 731,62
\bibitem[Ferland et~al.(1992)]{f92}
Ferland, G.~J., Peterson, B.~M., Horne, K., Welsh, W.~F., Nahar, S.~N., 1992, ApJ, 387, 95
\bibitem[\protect\citeauthoryear{Field}{1965}]{field65}
Field, G.~B., 1965, ApJ, 142, 531
\bibitem[\protect\citeauthoryear{Fryxell et al.}{2000}]{fr00}
Fryxell, B., Olson, K., Ricker, P., et al., 2000, ApJS, 131, 273
\bibitem[\protect\citeauthoryear{Gazol et al.}{2001}]{gaz01}
Gazol, A., V\'{a}zquez-Semadeni, E., S\'{a}nchez-Salcedo, F. J., Scalo, J., 2001, ApJ, 557, L121
\bibitem[\protect\citeauthoryear{Gazol et al.}{2005}]{gaz05}
Gazol, A., V\'{a}zquez-Semadeni, E., Kim, J., 2005, ApJ, 630, 911
\bibitem[Glover \& Clark(2012a)]{gc12}
Glover, S.~C.~O., Clark, P.~C., 2012a, MNRAS, 421, 9
\bibitem[Glover \& Clark(2012b)]{gc12b}
Glover, S.~C.~O., Clark, P.~C., 2012b, MNRAS, 426, 377
\bibitem[Glover et~al.(2010)]{gf10}
Glover, S. C. O., Federrath, C., Mac Low, M.-M., Klessen, R. S., 2010, MNRAS, 404, 2
\bibitem[\protect\citeauthoryear{Glover \& Mac Low}{2007a}]{gml07a}
Glover, S. C. O., Mac Low, M.-M., 2007a, ApJS, 169, 239
\bibitem[\protect\citeauthoryear{Glover \& Mac Low}{2007b}]{gml07b}
Glover, S. C. O., Mac Low, M.-M., 2007b, ApJ, 659, 1317
\bibitem[\protect\citeauthoryear{Gressel}{2009}]{gressel09}
Gressel, O., 2009, A\&A, 498, 661
\bibitem[Habing(1968)]{habing68}
Habing, H.~J., 1968, Bull.\ Astron.\ Inst.\ Netherlands, 19, 421
\bibitem[Hartmann et al.(2001)]{hartmannetal01} 
Hartmann, L., Ballesteros-Paredes, J., Bergin, E.~A.\ 2001, ApJ, 562, 852
\bibitem[Heitsch et al.(2005)]{heit05}
Heitsch, F., Burkert, A., Hartmann, L.~W., Slyz, A.~D., \& Devriendt, J.~E.~G., 2005, ApJ, 633, L113
\bibitem[\protect\citeauthoryear{Heitsch \& Hartmann}{2008}]{heitsch08}
Heitsch, F., Hartmann, L. 2008, ApJ, 689, 290
\bibitem[\protect\citeauthoryear{Hennebelle \& P\'{e}rault}{1999}]{hen99}
Hennebelle, P., P\'{e}rault, M., 1999, A\&A, 351, 309
\bibitem[\protect\citeauthoryear{Hennebelle et al.}{2008}]{hen08}
Hennebelle, P., Banerjee, R., V\'{a}zquez-Semadeni, E., Klessen, R. S.,Audit, E., 2008, A\&A, 486, L43
\bibitem[\protect\citeauthoryear{Heyer \& Brunt}{2004}]{hb04}
Heyer, M. H., Brunt, C. M., 2004, ApJ, 615, L45
\bibitem[\protect\citeauthoryear{Hollenbach \& McKee}{1979}]{hm79}
Hollenbach, D., McKee, C.~F., 1979, ApJS, 41, 555
\bibitem[Hughes et~al.(2010)]{hughes10} 
Hughes, A., et~al. 2010, MNRAS, 406, 2065
\bibitem[\protect\citeauthoryear{Koyama \& Inutsuka}{2000}]{ki00}
Koyama, H., Inutsuka, S.-I., 2000, ApJ, 532, 980
\bibitem[\protect\citeauthoryear{Koyama \& Inutsuka}{2002}]{ki02}
Koyama, H., Inutsuka, S.-I., 2002, ApJ, 564, L97
\bibitem[\protect\citeauthoryear{Larson}{1981}]{lar81}
Larson, R. B., 1981, MNRAS, 194, 809
\bibitem[{Le Teuff}, Millar, \& Markwick(2000)]{umist99}
{Le Teuff}, Y.~H., Millar, T.~J., \& Markwick, A.~J., 2000, A\&AS, 146, 157
\bibitem[{Mac Low} \& Shull(1986)]{ms86}
{Mac Low}, M.-M., Shull, J.~M., 1986, ApJ, 302, 585
\bibitem[\protect\citeauthoryear{Mac Low \& Klessen}{2004}]{rk04}
Mac Low, M.-M., Klessen, R. S., 2004, Rev. Mod. Phys., 76, 125
\bibitem[Martin, Keogh \& Mandy(1998)]{mkm98}
Martin, P.~G., Keogh W.~J., Mandy, M.~E., 1998, ApJ, 499, 793
\bibitem[McKee \& Ostriker(2007)]{mo07}
McKee, C.~F., \& Ostriker, E.~C.\ 2007, ARA\&A, 45, 565
\bibitem[\protect\citeauthoryear{Men'shchikov et al.}{2010}]{men10}
Men'shchikov, A., Andr\'{e}, Ph., Didelon, P., et al., 2010, A\&A, 518L, 103M
\bibitem[\protect\citeauthoryear{Micic et al.}{2012}]{mm12}
Micic, M., Glover, S. C. O, Federrath, C., Klessen, R. S., 2012, 421, 2531
\bibitem[\protect\citeauthoryear{Myers}{1983}]{my83}
Myers, P. C., 1983, ApJ, 270, 105
\bibitem[\protect\citeauthoryear{Powell et al.}{1999}]{powell99}
Powell, K.~G., Roe, P.~L., Linde, T.~J., Gombosi, T.~I., de Zeeuw, D.~L., 1999, JCoPh, 154, 284
\bibitem[Roman-Duval et~al.(2010)]{rd10}
Roman-Duval, J., Jackson, J.~M., Heyer, M., Rathborne, J., \& Simon, R.\ 2010, ApJ, 723, 492
\bibitem[\protect\citeauthoryear{Scalo \& Elmegreen}{2004}]{se04}
Scalo J., Elmegreen B. G., 2004, ARA\&A, 42, 275
\bibitem[Scoville \& Hersh(1979)]{sh79}
Scoville, N.~Z., \& Hersh, K.\ 1979, ApJ, 229, 578
\bibitem[\protect\citeauthoryear{Sembach et al.}{2000}]{s00}
Sembach, K. R., Howk, J. C., Ryans, R. S. I., Keenan, F. P., 2000, ApJ, 528, 310
\bibitem[\protect\citeauthoryear{Solomon et al.}{1987}]{sol87}
Solomon, P. M., Rivolo, A. R., Barrett, J., Yahil, A., 1987, ApJ, 319, 730
\bibitem[Trevisan \& Tennyson(2002)]{tt02}
Trevisan, C.~S., Tennyson, J., 2002, Plasma Phys.\ Controlled Fusion, 44, 1263
\bibitem[\protect\citeauthoryear{Truelove et al.}{1997}]{truelove97}
Truelove, J. K., Klein, R. I., McKee, C. F., Holliman, J. H., Howell, L. H., Greenough, J. A, 1997, ApJ, 489, L179
\bibitem[\protect\citeauthoryear{V\'{a}zquez-Semadeni, Ballesteros-Paredes \& Rodriguez}{1997}]{vbr97}
V\'{a}zquez-Semadeni, E., Ballesteros-Paredes, J., Rodriguez, L. F., 1997, ApJ, 474, 292
\bibitem[\protect\citeauthoryear{V\'{a}zquez-Semadeni et al.}{2000}]{vaz00}
V\'{a}zquez-Semadeni, E., Gazol, A.,, 2000, ApJ, 540, 271
\bibitem[\protect\citeauthoryear{V\'{a}zquez-Semadeni et al.}{2003}]{vaz03}
V\'{a}zquez-Semadeni, E., Ballesteros-Paredes, J., Klessen, R., 2003, ApJ, 585, L131
\bibitem[\protect\citeauthoryear{V\'{a}zquez-Semadeni et al.}{2006}]{vaz06}
V\'{a}zquez-Semadeni, E., Ryu, D., Passot, T., Gonz\'{a}lez, R. F., Gazol, A., 2006, ApJ, 643, 245
\bibitem[\protect\citeauthoryear{V\'{a}zquez-Semadeni et al.}{2007}]{vaz07}
V\'{a}zquez-Semadeni, E., G\'{o}mez, G. C., Jappsen, A. K., Ballesteros-Paredes, J., Gonz\'{a}les, R. F., Klessen, R. S., 2007, ApJ, 657, 870
\bibitem[\protect\citeauthoryear{V\'{a}zquez-Semadeni et al.}{2011}]{vaz11}
V\'{a}zquez-Semadeni, E., Banerjee, R., G\'{o}mez, G. C., Hennebelle, P., Duffin, D., Klessen, R. S., 2011, MNRAS, 414, 2511 
\bibitem[\protect\citeauthoryear{Waagan}{2009}]{waagan09}
Waagan, K., 2009, JCoPh, 228, 8609
\bibitem[\protect\citeauthoryear{Waagan, Federrath \& Klingenberg}{2011}]{wfk11}
Waagan, K., Federrath, C., Klingenberg, C., 2011, JCoPh, 230, 3331
\bibitem[Weingartner \& Draine(2001a)]{wd01a}
Weingartner, J.~C., Draine, B.~T., 2001a, ApJ, 563, 842
\bibitem[Weingartner \& Draine(2001b)]{wd01b}
Weingartner, J.~C., Draine, B.~T., 2001b, ApJS, 134, 263
\bibitem[\protect\citeauthoryear{Wolfire et al.}{1995}]{w95}
Wolfire, M. G., Hollenbach, D., McKee, C. F., Tielens, A. G. G. M., Bakes, E. L. O., 1995, ApJ, 443, 152
\bibitem[Wolfire et~al.(2003)]{w03}
Wolfire, M.~G., McKee, C.~F., Hollenbach, D., Tielens, A.~G.~G.~M., 2003, ApJ, 587, 278
\bibitem[Woodward(1978)]{wood78}
Woodward, P.~R.\ 1978, ARA\&A, 16, 555
\bibitem[Zuckerman \& Palmer(1974)]{zp74}
Zuckerman, B., \& Palmer, P.\ 1974, ARA\&A, 12, 279

\end{thebibliography}
\end{document}